\documentclass[12pt]{article}
\usepackage[utf8]{inputenc}

\newcommand{\blind}{1}

\usepackage{amsgen,amsmath,amstext,amsbsy,amsopn,amssymb,mathabx,mathrsfs,amsthm,bm,bbm,romannum,enumitem,xr,multirow}
\usepackage{xstring}
\usepackage{caption}
\usepackage[dvips]{graphicx}
\usepackage{subcaption}
\usepackage[ruled,vlined,linesnumbered,resetcount]{algorithm2e}
\usepackage{setspace}
\usepackage[maxcitenames=1, maxbibnames=99,backend=biber,eprint=false,isbn=false,doi=false,url=false,citestyle=numeric-comp,giveninits=true]{biblatex}
\AtEveryBibitem{%
  \clearfield{month}%
}

\usepackage{hyperref}
\usepackage{cleveref}
\hypersetup{
	colorlinks=true,
	linkcolor=red,
	urlcolor=blue,
	citecolor=blue
}
\usepackage[depth=2]{bookmark}

\usepackage[margin=1in]{geometry}

\usepackage[utf8]{inputenc}

\usepackage{chngcntr}
\counterwithin{figure}{section}
\counterwithin{table}{section}

\newcommand{\new}{\color{black}}

\usepackage{booktabs}
\usepackage{longtable}
\usepackage{array}
\usepackage{multirow}
\usepackage[table]{xcolor}
\usepackage{wrapfig}
\usepackage{float}
\usepackage{colortbl}
\usepackage{pdflscape}
\usepackage{tabu}
\usepackage{threeparttable}
\usepackage{threeparttablex}
\usepackage[normalem]{ulem}
\usepackage{makecell}
\usepackage{pifont}
\newtheorem{thm}{Theorem}

\newtheorem{prop}[thm]{Proposition}
\newtheorem{coro}[thm]{Corollary}

\theoremstyle{definition}

\newtheorem{cond}{Condition}
\newtheorem{asm}{Assumption}

\theoremstyle{remark}

\newcommand{\bx}{\bm{x}}

\newcommand{\Ib}{\mathbf{I}}

\newcommand{\bX}{\bm{X}}

\newcommand{\bbE}{\mathbb{E}}

\newcommand{\bbP}{\mathbb{P}}
\newcommand{\bbQ}{\mathbb{Q}}
\newcommand{\bbR}{\mathbb{R}}

\newcommand{\cA}{\mathcal{A}}
\newcommand{\cB}{\mathcal{B}}

\newcommand{\cD}{\mathcal{D}}
\newcommand{\cE}{\mathcal{E}}

\newcommand{\cH}{\mathcal{H}}
\newcommand{\cI}{\mathcal{I}}

\newcommand{\cK}{\mathcal{K}}
\newcommand{\cL}{\mathcal{L}}

\newcommand{\cN}{\mathcal{N}}
\newcommand{\cO}{\mathcal{O}}
\newcommand{\cP}{\mathcal{P}}

\newcommand{\cR}{\mathcal{R}}

\newcommand{\sR}{\mathscr{R}}

\newcommand{\fC}{\mathfrak{C}}

\newcommand{\fM}{\mathfrak{M}}

\newcommand{\fR}{\mathfrak{R}}

\newcommand{\fV}{\mathfrak{V}}

\newcommand{\sfA}{\mathsf{A}}
\newcommand{\sfB}{\mathsf{B}}

\newcommand{\sfD}{\mathsf{D}}

\newcommand{\sfG}{\mathsf{G}}
\newcommand{\sfH}{\mathsf{H}}

\newcommand{\sfW}{\mathsf{W}}

\newcommand{\argmin}{\mathop{\mathrm{argmin}}}
\newcommand{\argmax}{\mathop{\mathrm{argmax}}}

\newcommand{\vvar}{\mathsf{Var}}

\newcommand{\cov}{\mathsf{Cov}}

\newcommand\indep{\protect\mathpalette{\protect\independenT}{\perp}}
\def\independenT#1#2{\mathrel{\rlap{$#1#2$}\mkern2mu{#1#2}}}	

\def\ubar#1{\underbar{$ #1 $}}

\newcommand{\rd}{\mathrm{d}}

\newcommand{\supp}{\mathsf{support}}

\newcommand{\train}{\mathrm{train}}
\newcommand{\test}{\mathrm{test}}
\newcommand{\post}{\mathrm{post}}
\newcommand{\ante}{\mathrm{ante}}
\newcommand{\emp}{\mathrm{emp}}
\newcommand{\IID}{\mathrm{IID}}
\newcommand{\conv}{\mathsf{conv}}

\usepackage{tikz}
\usetikzlibrary{calc}
\usetikzlibrary{positioning}
\usetikzlibrary{shapes,arrows}

\tikzstyle{do} = [draw, ellipse, text centered, 
text width = 1cm, minimum height = 2em]
\tikzstyle{content} = [draw, rectangle, rounded corners, text centered, 
fill = blue!20,  text width = 4cm, minimum height = 2em]
\tikzstyle{signal} = [draw, ellipse, text centered, 
fill = green!20, text width = 2cm, minimum height = 2em]
\tikzstyle{payoff} = [draw, ellipse, text centered,
fill = red!20, text width = 2cm, minimum height = 2em]
\tikzstyle{action} = [draw, diamond, aspect = 2, text centered, 
fill = yellow!20, text width = 2cm, minimum height = 2em]

\begin{document}

\pagenumbering{arabic}

\def\spacingset#1{\renewcommand{\baselinestretch}%
{#1}\small\normalsize} \spacingset{1}


\if1\blind
{
  \title{\bf     \vspace{-15mm} Minimax Regret Learning for Data with Heterogeneous Sub-populations\vspace{-2mm}}
  \author{
    Weibin Mo
    \and
    Weijing Tang
    \and
    Songkai Xue
    \and
    Yufeng Liu
    \and 
    Ji Zhu%
    \thanks{
        Weibin Mo and Weijing Tang are co-first authors and have made equal contributions.
        Weibin Mo is Assistant Professor in the Daniels School of Business at Purdue University. 
        Weijing Tang is Assistant Professor in the Department of Statistics and Data Science at Carnegie Mellon University.
        Songkai Xue is a Ph.D. student in the Department of Statistics at the University of Michigan.
        Yufeng Liu 
        and Ji Zhu are Professors in the Department of Statistics at the University of Michigan.
    }
  }
    \date{}
  \maketitle
  \vspace{-15mm}
} \fi

\if0\blind
{
  \bigskip
  \bigskip
  \bigskip
  \begin{center}
    {\LARGE\bf Minimax Regret Learning for Data with Heterogeneous Sub-populations}
\end{center}
  \medskip
} \fi

\bigskip
\begin{abstract}
Modern complex datasets often consist of various sub-populations with known group information.
In the presence of sub-population heterogeneity, 
it is crucial to develop robust and generalizable learning methods that 
(1) can enjoy robust performance on each of the training populations, and 
(2) is generalizable to an unseen testing population.
While various min-max formulations have been proposed to achieve (1) in the robust learning literature, their generalization to an unseen testing is less explored.
Moreover, a general min-max formulation can be sensitive to the noise heterogeneity, and, in the extreme case, be degenerate such that a single high-noise population dominates \parencite{agarwal2022minimax}.  
The min-max-regret (MMR) can mitigate these challenges.
In this work, we consider a distribution-free robust hierarchical model for the generalization from multiple training populations to an unseen testing population.
Under the robust hierarchical model,
the empirical MMR can enjoy the regret guarantees on each of the training populations as well as the unseen testing population.
We further specialize the general MMR framework to linear regression and generalized linear model, 
where we characterize the geometry of MMR and its distinction from other robust methods.
We demonstrate the effectiveness of MMR through extensive simulation studies and an application to image recognition.
\end{abstract}

\noindent%
{\it Keywords:} 
Generalizability;
Heterogeneous sub-populations; 
Hierarchical model;
Meta analysis; 
Minimax regret; 
Robust learning.

\spacingset{1.7} 
\setlength{\abovedisplayskip}{2pt}
\setlength{\belowdisplayskip}{2pt}

\vspace{-7mm}
\section{Introduction}
\label{Sec: intro}

In modern big data era, complex datasets in various fields often consist of heterogeneous sub-populations, such as different demographics or socioeconomic statuses in health disparities~\parencite{national2017state}, various cell types in gene expression~\parencite{li2019anchor}, or diverse domains in natural language processing~\parencite{cer-etal-2017-semeval}. 
Such sub-populations can correspond to heterogeneous covariate distributions, covariate-response relationships, as well as 
heterogeneous goodness of model fits at the sub-population level.
Due to substantial heterogeneity across sub-populations, predictive models that optimize the average performance over the pooled population may suffer from poor generalizability to certain sub-populations~\parencite{blodgett-etal-2016-demographic,tatman-2017-gender,zech2018variable}.
It is crucial to develop robust and generalizable statistical learning methods for high-stakes and fairness-critical decision making such as medical diagnosis and criminal justice, 
which can enjoy uniformly good performance across heterogeneous training populations, 
and can be generalizable to an unseen testing population that is different from training. 

{\new
We consider the setting with a meta dataset consisting of $K$ different samples, and each sample is a dataset from a distinct training population.
It is equivalent to the data with $K$ known groups, and the group information of each data point is available. 
For example, the electronic health record (EHR) data collected from multiple hospitals over various time periods can be grouped based on their sources~\parencite{nestor2019feature, gola2020population, singh2022generalizability}. 
More meta-data applications with known groups for pattern recognition and natural language processing can be found in \textcite{sagawa2019distributionally} and \textcite{koh2021wilds}. 
In these applications, we aim for a generic learning methodology that can enjoy robust performance on each of the training populations, and is also generalizable to an unseen testing population.

In the existing literature, the robust generalization to multiple populations $\bbP^{(1)},\cdots,\bbP^{(K)}$ can be achieved via the min-max optimization \parencite{meinshausen2015maximin,sagawa2019distributionally,agarwal2022minimax}:
\begin{align}
    \min_{\theta\in\Theta}\max_{1 \le k \le K}F(\theta,\bbP^{(k)}),
    \label{eq:min-max}
\end{align}
where $F(\theta,\bbP)$ is a smaller-the-better criterion function of the parameter of interest $\theta$ and the underlying population $\bbP$.
Such a min-max formulation can guarantee the generalization performance on each of the training populations $\bbP^{(1)},\cdots,\bbP^{(K)}$ where we have access to data,
but its generalization to an unseen testing population $\bbP_{\test}$ where we have no access to data remains unclear.
Moreover, the criterion function $F(\theta,\bbP)$ plays an important role in the solution to \eqref{eq:min-max}.
While the risk function can be a natural choice of $F(\theta,\bbP)$ \parencite{sagawa2019distributionally}, it is known that the corresponding min-max solution can be sensitive to the noise heterogeneity \parencite{agarwal2022minimax}. 
In the extreme case, the min-max problem \eqref{eq:min-max} can degenerate to $\min_{\theta\in\Theta}F(\theta,\bbP^{(k^{\star})})$, which is the risk minimization on a single high-noise population $\bbP^{(k^{\star})}$.
These challenges can be mitigated if the regret function is considered for $F(\theta,\bbP)$ \parencite{agarwal2022minimax}.
Other alternatives such as the risk difference against a reference \parencite{meinshausen2015maximin} may not achieve this goal,
and their sensitivity to nuisance heterogeneity and degeneration are not fully characterized in the related literature.

In this paper, we consider a distribution-free robust hierarchical model for the generalization from multiple heterogeneous training populations to an unseen testing population.\footnote{
    This is different from the setting in transfer learning \parencite{li2022transfer,li2024estimation,cai2024transfer}, where data from the testing population are available, and model-based assumptions are imposed among $\bbP^{(1)},\cdots,\bbP^{(K)},\bbP_{\test}$.
    There is a parallel line of research that aims to recover an invariant prediction model across multiple heterogeneous populations, which can be applied to an unseen testing population \parencite{peters2016causal,arjovsky2019invariant,fan2024environment,gu2024causality}. These works rely on certain invariance structures among $\bbP^{(1)},\cdots,\bbP^{(K)},\bbP_{\test}$, which are different from our robust hierarchical model.
}
It is motivated from the meta-analysis literature \parencite{hedges1985statistical,brantner2023methods} and the recent advancements in conformal inference across multiple populations \parencite{dunn2023distribution,andrews2022transfer}.
The populations and data are randomly generated at two levels.
At the meta-level, we consider some training meta-distribution $\fM_{\train}$ that generates $\bbP^{(1)},\cdots,\bbP^{(K)}$ independently and 
some testing meta-distribution $\fM_{\test}$ that generates $\bbP_{\test}$,
while allowing that $\fM_{\train}\neq\fM_{\test}$ for generalization robustness.
At the data-level, the observed $K$ samples are generated from $\bbP^{(1)},\cdots,\bbP^{(K)}$ independently.
We impose no distributional assumptions on the meta- and data-distributions,
with the only restriction that $\bbP_{\test}$ is realizable from $\fM_{\train}$ at the meta-level.

Under the robust hierarchical model, we propose two nested generalization criteria: 
the ex-post worst-case regret among $\{\bbP^{(k)}\}_{k=1}^{K}$, 
and the ex-ante worst-case regret among $\cP$, 
where $\cP$ consists of all realizable populations from $\fM_{\train}$.
The ex-post criterion is a performance guarantee for the training populations $\{\bbP^{(k)}\}_{k=1}^{K}$,
which has been widely considered in the robust learning literature \parencite{meinshausen2015maximin,sagawa2019distributionally,agarwal2022minimax}.
Our ex-ante criterion is a performance guarantee for the unseen testing population $\bbP_{\test}$,
which, to the best of our knowledge, is the first generalization guarantee for an unseen testing under the distribution-free robust hierarchical model.

Given the generalization criteria, our learning goal is to minimize both the ex-post and ex-ante worst-case regrets, which we refer to as the \textbf{ex-post} and \textbf{ex-ante min-max-regret (MMR)} problems.
In particular, the ex-post MMR solves \eqref{eq:min-max} with $F(\theta,\bbP)$ being the regret function.
To distinguish MMR from other robust methods based on \eqref{eq:min-max}, 
we further characterize the duality of \eqref{eq:min-max}, which entails its sensitivity to nuisance heterogeneity and potential degeneration whenever $F(\theta,\bbP)$ is not a regret function.
Based on the training samples, 
we further consider the \textbf{empirical MMR} as an empirical analog of the ex-post MMR, 
and establish its ex-post and ex-ante MMR guarantees.
In this way, the empirical MMR can enjoy the regret guarantees on each of the training populations as well as the unseen testing population under the robust hierarchical model.
}

The rest of this paper is organized as follows. 
We introduce the {\new robust hierarchical model} and the MMR framework in Section~\ref{sec:MMR}. 
Our MMR framework is considered for a general learning problem with a given loss function, and does not rely on model assumptions. 
We further consider its specialization to the linear regression (Section \ref{sec:LM}) and 
{\new generalized linear model (GLM) (Section \ref{sec:GLM})} settings, 
where we shed further lights on its properties and the comparisons with existing robust methods.
In Section~\ref{sec:theory}, we establish the ex-post and ex-ante MMR guarantees for the empirical MMR. 
We validate our findings through extensive simulation studies in Section~\ref{sec:simulation} and a real-world application to {\new image recognition} in Section~\ref{sec:realdata}.

{\new
\vspace{-7mm}
\section{The Min-Max-Regret (MMR) Framework}\label{sec:MMR}
\vspace{-2mm}
We consider a general learning problem with multiple heterogeneous training populations and an unseen testing population.
The learning task is based on a given loss function $\ell_{\theta}(Z)$, which is a smaller-the-better function of the data $Z$ and the parameter of interest $\theta$, and a given parameter space $\Theta$.
This leads to the risk function $R^{\dagger}(\theta,\bbP) = \bbE_{\bbP}[\ell_{\theta}(Z)]$,  where the expectation is taken with respect to the data $Z$ under the distribution $\bbP$.
During the training stage, we observe training samples from $K$ populations: 
$\cD^{(k)} = \{ Z_{i}^{(k)} \}_{i=1}^{n_{k}} \overset{\IID}{\sim} \bbP^{(k)}$, 
where $n_{k}$ is the sample size, 
and $\bbP^{(k)}$ is the $k$-th data generation distribution,
for $k=1,\cdots,K$.
The training populations $\bbP^{(1)},\cdots,\bbP^{(K)}$ are considered generally heterogeneous.
During the testing stage, the population of interest is a testing distribution $\bbP_{\test}$, which can be different from any of the training distributions $\bbP^{(1)},\cdots,\bbP^{(K)}$.
The learning goal is to obtain an estimate $\widehat{\theta}$ from $\Theta$ based on the training samples $\cD^{(1)},\cdots,\cD^{(K)}$, such that the generalization error on $\bbP_{\test}$, measured as the \textit{regret}, is smaller the better:
\begin{align}
	R(\widehat{\theta},\bbP_{\test}) := R^{\dagger}(\widehat{\theta},\bbP_{\test}) - \underbrace{\inf_{\beta\in\Theta}R^{\dagger}(\beta,\bbP_{\test})}_{\text{WMR}~\fR^{\dagger}(\bbP_{\test})}.
	\label{eq:regret}
\end{align}
In particular, the regret compares the risk of $\widehat{\theta}$ with the \textit{within-population minimized risk (WMR)} on $\bbP_{\test}$. 
The WMR $\fR^{\dagger}(\bbP) := \inf_{\beta\in\Theta}R^{\dagger}(\beta,\bbP)$ is the lowest achievable risk on a data distribution $\bbP$ among $\theta\in\Theta$.
\footnote{
If $\Theta$ is an unrestricted class of measurable functions of data, then $\fR^{\dagger}(\bbP)$ 
is often referred to as the \textit{Bayes risk} 
on the population $\bbP$.
}

In the following Section \ref{sec:generalization}, we introduce the robust hierarchical model for the generalization from training to testing,
which leads to the ex-post and ex-ante MMR criteria.
Then we introduce the empirical MMR as a learning methodology based on data in Section \ref{sec:empirical_MMR}.
The comparisons of our MMR with the existing literature are provided in Section \ref{sec:compare}. 
The optimization algorithm for empirical MMR is further discussed in Section \ref{sec:algo}.

\vspace{-7mm}
\subsection{Robust Hierarchical Model}\label{sec:generalization}
\vspace{-2mm}
The generalization from training to testing requires further assumptions on the relationships among $\{\bbP^{(1)},\cdots,\bbP^{(K)},\bbP_{\test}\}$.
In particular, we aim to allow that $\bbP_{\test}$ is unseen, in the sense that it is not directly represented by the observed training distributions $\bbP^{(1)},\cdots,\bbP^{(K)}$.

We adopt the following hierarchical model on the relationships among the data and the training and testing distributions.
During the training stage, there is a fixed but unknown training \textbf{meta-distribution} $\fM_{\train}$ as a probability measure on the space of data distributions, 
such that the training data distributions $\bbP^{(1)},\cdots,\bbP^{(K)}$ are independently generated from $\fM_{\train}$.
Conditional on the realization of $\bbP^{(1)},\cdots,\bbP^{(K)}$, the samples $\cD^{(1)},\cdots,\cD^{(K)}$ are further generated independently from $\bbP^{(1)},\cdots,\bbP^{(K)}$, respectively.
During the testing stage, the testing data distribution $\bbP_{\test}$ is generated in an analogous manner, but from another fixed but unknown testing meta-distribution $\fM_{\test}$. 
For robust generalization, we allow the training and testing meta-distributions, $\fM_{\train}$ and $\fM_{\test}$, to differ, with the requirement that $\cP := \supp(\fM_{\train}) \supseteq \supp(\fM_{\test})$.
Here, $\cP$ encloses all ex-ante realizable data-distributions from training. Without further assumptions, the condition $\supp(\fM_{\test})\subseteq\cP$ implies that $\bbP_{\test}$ can be any data distribution in $\cP$ that is ex-ante realizable during training.
These relationships are illustrated in Figure \ref{fig:generlization}.

\begin{figure}[!t]
	\centering
	\resizebox{0.8\textwidth}{!}{
        \begin{tikzpicture}[->]
			\node (P) [draw, ellipse, text centered] 
			{$\cP$};
			
			\node [left = 1cm of P, color = gray] 
			{(\textit{Unobserved})};
			\node [draw, dashed, text centered, left = 4cm of P] (ex-ante)
			{\makecell{
					\textit{Ex-ante realizable} \\
					\textit{data-distributions}
			}};
			
			\node [draw, ellipse, text centered, below left = 1cm and 2cm of P] (P1) {$ \bbP^{(1)} $};
			\node [draw, ellipse, text centered, below right = 1cm and 2cm of P] (PK) {$ \bbP^{(K)} $};
			\node [text centered, below = 1cm of P] (P-dots) {$ \cdots $};
			\node [draw, ellipse, text centered, color = red, below right = 1cm and 5cm of P] (P-test) {$ \bbP_{\test} $};
			\path 
			(P) edge (P1)
			(P) edge 
			node [right] {$\fM_{\train}$} 
			(P-dots)
			(P) edge (PK)
			(P.east) edge [color = red] 
			node [above right] {$\fM_{\test}$} 
			(P-test);
			
			\node [draw, dashed, text centered, below = .5cm of ex-ante] (ex-post)
			{\makecell{
					\textit{Ex-post realized} \\
					\textit{data-distributions}
			}};
			
			\draw [dashed]
			($ (P.north west) + (-8, 1) $)
			rectangle 
			($ (P-test.south east) + (1, -.75) $);
			
			\node [text centered, below = 1cm of P1] (D1) {$ \{ Z_{i}^{(1)} \}_{i=1}^{n_{1}} $};
			\node [text centered, below = 1cm of PK] (DK) {$ \{ Z_{i}^{(K)} \}_{i=1}^{n_{K}} $};
			\node [text centered, below = 1.5cm of P-dots] (D-dots) {$ \cdots $};
			\path 
			(P1) edge (D1)
			(P-dots) edge (D-dots)
			(PK) edge (DK);
			
			\node [draw, dashed, text centered, below = 1cm of ex-post] (data) 
			{\textit{Data}};
			
			\node [draw, ellipse, text centered, color = blue, text width = .5cm, below = 1cm of D-dots] (estim) {$ \widehat{\theta} $};
			\path 
			(D1) edge (estim)
			(D-dots) edge (estim)
			(DK) edge (estim);
			\node [draw, dashed, text centered, color = blue, below = 1cm of data] (Estim) {\textit{Estimator}};
			
			\node [draw, dashed, text centered, color = violet, below = 1cm of Estim] (eval) {\textit{Evaluation criteria}};
			
			\node [draw, text centered, color = violet, below = .5cm of estim] (ex-post-MMR) {\makecell{
					\textbf{Ex-post MMR:}
					worst-case \\ 
					regret among $\{\bbP^{(k)}\}_{k=1}^{K}$
			}};
			\path 
			(P1) edge [color = violet, dotted, line width = 1, bend right = 60] (ex-post-MMR)
			(P-dots) edge [color = violet, dotted, line width = 1, bend right = 30] (ex-post-MMR)
			(PK) edge [color = violet, dotted, line width = 1, bend left = 60] (ex-post-MMR);
			
			\node [draw, text centered, color = violet, right = 1cm of ex-post-MMR] (ex-ante-MMR) {\makecell{
					\textbf{Ex-ante MMR:} worst- \\
					case regret among $\cP$
			}};
			\draw [color = red, dotted, line width = 1]
			(P-test.south) -- (P-test|-ex-ante-MMR.north);
			\path 
			(estim) edge [color = blue] (ex-post-MMR)
			(estim) edge [bend left = 10, color = blue] (ex-ante-MMR);
		\end{tikzpicture}
    }
	\caption{
		Generalization of heterogeneous training data distributions $\{\bbP^{(k)}\}_{k=1}^{K}$ to testing $\bbP_{\test}$,
		where $\fM_{\train},\fM_{\test}$ are meta-distributions that generate data distributions from $\cP$.
		It incorporates the ex-post perspective, where $\bbP^{(1)},\cdots,\bbP^{(K)}$ given as fixed, 
        and the ex-ante perspective, where $\bbP^{(1)},\cdots,\bbP^{(K)}\overset{\rm IID}{\sim} \fM_{\train}$ and $\bbP_{\test} \sim \fM_{\test}$ with meta-distributions satisfying $\cP=\mathsf{support}(\fM_{\train}) \supseteq \mathsf{support}(\fM_{\test})$. 
        }
	\label{fig:generlization}
    \vspace{-4mm}
\end{figure}

The hierarchical model for heterogeneous data distributions has been widely used in meta-analysis \parencite{hedges1985statistical,brantner2023methods},
although it is often considered under the parametric settings, and referred to as the random effect model \parencite{laird1982random}.
Different from the traditional meta-analysis, 
we (1) impose no distributional assumptions on the meta-distributions $\fM_{\train},\fM_{\test}$ and data-distributions $\bbP^{(1)},\cdots,\bbP^{(K)},\bbP_{\test}$, and (2) consider $\fM_{\train}$ and $\fM_{\test}$ as generally distinct.
For (1), to serve a general learning problem, our generalization framework is distribution-free and only relies on the loss function instead of a pre-specified parametric model.
Such a distribution-free hierarchical model has also been studied in \parencite{dunn2023distribution,andrews2022transfer} recently for conformal inference across multiple populations.
For (2), we aim for a robust generalizability guarantee to allow systematic discrepancies between training and testing.
\footnote{
	In the conformal inference literature \parencite{dunn2023distribution,andrews2022transfer},
	$\fM_{\train}=\fM_{\test}$ (exchangeability) is often assumed.
	While \parencite{andrews2022transfer} allow $\fM_{\train}\neq\fM_{\test}$, they have assumed additional conditions for 
    $\rd\fM_{\test}/\rd\fM_{\train}$ to perform weighted conformal inference.
	Our paper aims for a different goal, 
	and does not rely on such conditions.
}

The purpose of introducing a hierarchical model is to distinguish two types of generalizability from the \textbf{ex-post}  and \textbf{ex-ante} perspectives \parencite{dunn2023distribution,andrews2022transfer}.
From the ex-post perspective that $\bbP^{(1)},\cdots,\bbP^{(K)}$ are given as fixed, we consider that the testing task is carried out on one of these populations,
that is, to assume that $\bbP_{\test}\in\{\bbP^{(1)},\cdots,\bbP^{(K)}\}$.
This is a common underlying assumption in the robust learning literature with multiple populations \parencite{sagawa2019distributionally,agarwal2022minimax,guo2023statistical}.
In this setting, a valid generalizability guarantee is the worst-case regret among $\bbP^{(1)},\cdots,\bbP^{(K)}$:
\begin{align}
	\cR_{\post}(\theta) := \max_{1 \le k \le K}R(\theta,\bbP^{(k)}).
	\label{eq:regret_post}
\end{align}
We refer to the objective that minimizes such an ex-post worst-case regret as the \textbf{ex-post MMR}.
Note that $\cR_{\post}(\theta)$ is also the worst-case regret for $\bbP_{\test} \in \big\{ \sum_{k=1}^{K}\gamma_{k}\bbP^{(k)}: \gamma\in\Delta^{K-1} \big\}$.\footnote{
    It follows from Supplementary Material Lemma H.1.
    Here, we denote
    $\Delta^{K-1} := \{ (\gamma_{1},\cdots,\gamma_{K})^{\intercal}: \gamma_{1},\cdots,\gamma_{K} \ge 0, \sum_{k=1}^{K}\gamma_{k} = 1 \}$ 
    as the $(K-1)$-dimensional simplex.
}

From the ex-ante perspective, the data distributions $\bbP^{(1)},\cdots,\bbP^{(K)},\bbP_{\test}$ are considered random realizations from the respective meta-distributions $\fM_{\train},\fM_{\test}$. In particular,
we consider $\fM_{\train}$, that generates $\bbP^{(1)},\cdots,\bbP^{(K)}$, and $\fM_{\test}$, that generates $\bbP_{\test}$, are supported on the same unobserved $\cP$.
In this way, the testing data-distribution $\bbP_{\test}$ needs not be representable as a convex combination of the training ones $\{\bbP^{(k)}\}_{k=1}^{K}$,
that is, $\bbP_{\test}\neq\sum_{k=1}^{K}\gamma_{k}\bbP^{(k)}$ for any $\gamma\in\Delta^{K-1}$.
Instead, they are connected via the unobserved $\cP$ for ex-ante realizable data distributions. 
In this setting, we consider the worst-case regret among $\cP$:
\begin{align}
	\cR_{\ante}(\theta) := \sup_{\bbP\in\cP}R(\theta,\bbP)
	\label{eq:regret_ante}
\end{align}
as the ex-ante generalizability guarantee.
We refer to the objective that minimizes such an ex-ante worst-case regret as the \textbf{ex-ante MMR}.

By definition, $\{\bbP^{(k)}\}_{k=1}^{K}\subseteq\cP$, and hence the ex-ante MMR \eqref{eq:regret_ante} is an upper bound of the ex-post MMR \eqref{eq:regret_post}.
The additional ex-ante robustness is due to the generalization to an unseen testing beyond the training realizations.
The choice of the generalization guarantee between ex-post and ex-ante depends on the practical needs. If the generalization is to all training populations and their mixtures, then the ex-post MMR is sufficient and less conservative. 
If the generalization is beyond such mixtures, then the ex-ante MMR is applicable.
}

\vspace{-7mm}
\subsection{Empirical MMR}\label{sec:empirical_MMR}
\vspace{-2mm}
Despite that the ex-post and ex-ante MMR criteria have different generalization scopes and interpretations, we consider a single learning methodology based on the observed samples from $K$ populations: $\cD^{(k)}=\{Z_{i}^{(k)}\}_{k=1}^{n_{k}}$ for $k=1,\cdots,K$.
Specifically, we formulate our empirical goal as to solve the \textbf{empirical MMR} problem:
\begin{align}
	\min_{\theta\in\Theta}\max_{1 \le k \le K}\left\{ R\big( \theta,\bbP_{n_{k}}^{(k)} \big) := {1 \over n_{k}}\sum_{i=1}^{n_{k}}\ell_{\theta}(Z_{i}^{(k)}) - \inf_{\beta \in \Theta}{1 \over n_{k}}\sum_{i=1}^{n_{k}}\ell_{\beta}(Z_{i}^{(k)}) \right\}.
	\label{eq:MMR}
\end{align}
Here, the inner-most minimization $\inf_{\beta \in \Theta}{1 \over n_{k}}\sum_{i=1}^{n_{k}}\ell_{\beta}(Z_{i}^{(k)})$ is a within-sample empirical risk minimization (ERM) problem, which estimates the WMR $\fR^{\dagger}(\bbP^{(k)})$.
The min-max objective $R\big( \theta,\bbP_{n_{k}}^{(k)} \big)$ is the \textit{empirical regret}, where $\bbP_{n_{k}}^{(k)}$ is the empirical distribution based on the samples $\cD^{(k)}$,
and $R\big( \theta,\bbP_{n_{k}}^{(k)} \big)$ is an empirical analog of the regret function $R(\theta,\bbP^{(k)})$ on the $k$-th training data distribution $\bbP^{(k)}$.
The middle layer of maximization in~\eqref{eq:MMR} takes the worst case among $K$ populations,
while the outer minimization obtains the empirical MMR estimate $\widehat{\theta}$. 
{\new
The empirical MMR can achieve both the ex-post and ex-ante MMR guarantees. 
Specifically, consider the worst-case empirical regret among the observed $K$ samples:
\begin{align}
	\cR_{\emp}(\theta) := \max_{1 \le k \le K}R\big( \theta, \bbP_{n_{k}}^{(k)} \big).
	\label{eq:regret_emp}
\end{align}
Then $\widehat{\theta}\in\argmin_{\theta\in\Theta}\cR_{\emp}(\theta)$, which can also minimize $\cR_{\post}(\theta)$ and $\cR_{\ante}(\theta)$ due to the approximation:
$\cR_{\emp}(\theta) \approx \cR_{\post}(\theta) \approx \cR_{\ante}(\theta)$.
The first approximation is to use the empirical distributions $\bbP_{n_{1}}^{(1)},\cdots,\bbP_{n_{K}}^{(K)}$ to approximate the training data distributions $\bbP^{(1)},\cdots,\bbP^{(K)}$.
The second one is to use the maximum regret among $\{\bbP^{(k)}\}_{k=1}^{K}$ to approximate the supremum regret among $\cP$.
Formal theoretical results are established in Section \ref{sec:theory}.
}

{\new
\vspace{-7mm}
\subsection{Relationships with Existing Methods}\label{sec:compare}
\vspace{-2mm}
\paragraph{Pooled ERM}
Given the samples $\cD^{(1)},\cdots,\cD^{(K)}$ from potentially heterogeneous populations, 
one may overlook their heterogeneity and consider the \textit{pooled ERM} problem:
\begin{align}
	\min_{\theta \in \Theta}{1 \over \sum_{k=1}^{K}n_{k}}\sum_{k=1}^{K}\sum_{i=1}^{n_{k}}\ell_{\theta}(Z_{i}^{(k)}).
	\label{eq:ERM_pooled}
\end{align}
This effectively minimizes the risk on a particular mixture of training populations $\sum_{k=1}^{K}\gamma_{k}\bbP^{(k)}$, where $\gamma_{k} = n_{k}/\sum_{k'=1}^{K}n_{k'}$.
Such a mixture explicitly depends on the relative sample sizes of $n_{1},\cdots,n_{K}$, and can be susceptible to their variations. 
Moreover, it optimizes the performance on a weighted average of the training populations $\{\bbP^{(k)}\}_{k=1}^{K}$.
When $\{\bbP^{(k)}\}_{k=1}^{K}$ are heterogeneous, 
the weighted average is not robust to generalize to each of $\bbP^{(1)},\cdots,\bbP^{(K)}$, and can suffer from poor performance on some of these training populations.
From the meta-perspective in Section \ref{sec:generalization}, the weighted average among the ex-post training populations $\{\bbP^{(k)}\}_{k=1}^{K}$ depends on the underlying training meta-distribution $\fM_{\train}$, and is not robust to generalize to testing when $\fM_{\train}\neq\fM_{\test}$.
More discussions on its non-robustness are provided in Supplementary Material B.

\vspace{-5mm}
\paragraph{Group Distributionally Robust Optimization (GDRO)}
To mitigate the non-robustness of pooled ERM, 
the \textit{group distributionally robust optimization (GDRO)} \parencite{hu2018does,sagawa2019distributionally} was proposed to solve the min-max risk problem:
\begin{align}
	\min_{\theta\in\Theta}\max_{1 \le k \le K}R^{\dagger}(\theta,\bbP^{(k)}).
	\label{eq:GDRO}\end{align}
It can be equivalently considered to minimize the worst-case risk among $\bbP_{\test}\in\big\{\sum_{k=1}^{K}\gamma_{k}\bbP^{(k)}:\gamma\in\Delta^{K-1}\big\}$.
Compared to our ex-post MMR \eqref{eq:regret_post},
GDRO is based on the risk function $R^{\dagger}(\theta,\bbP^{(k)})$ without subtracting the WMR $\fR^{\dagger}(\bbP^{(k)})=\inf_{\beta\in\Theta}R^{\dagger}(\beta,\bbP^{(k)})$.
In terms of generalization, 
GDRO and MMR can both enjoy robust generalization guarantees, but are different in the criteria to measure generalization. 
In particular, the generalizations of GDRO and MMR are measured by the testing risk $R^{\dagger}(\theta,\bbP_{\test})$ and regret $R(\theta,\bbP_{\test})$, respectively.

In practice, the training WMRs $\{\fR^{\dagger}(\bbP^{(k)})\}_{k=1}^{K}$ can be heterogeneous,
which is possibly due to the varying qualities of training samples, or the incorporation of noisy data.
In these cases, GDRO can be sensitive to the high WMRs of certain noisy populations, 
and can even degenerate to their risk minimizers that are conservative and uninformative.
In contrast, MMR avoids such a challenge by subtracting the WMR in its criterion function.
In Supplementary Material C,
we show that the dual GDRO problem is
$\max\big\{\fR^{\dagger}(\bbQ):\bbQ=\sum_{k=1}^{K}\gamma_{k}\bbP^{(k)},\gamma\in\Delta^{K-1}\big\}$,
which explains its sensitivity to the heterogeneity among $\{\fR^{\dagger}(\bbP^{(k)})\}_{k=1}^{K}$.
We further characterize the degeneration that 
GDRO is dominated by a single training population when the corresponding WMR dominates,
while MMR degenerates only when it achieves zero regrets simultaneously across $\{\bbP^{(k)}\}_{k=1}^{K}$.
In Section \ref{sec:compare_LM}, we have a detailed comparison of the GDRO and MMR in terms of their sensitivity to the heterogeneous WMRs in linear regression. 
More numerical comparisons are provided in Section \ref{sec:simulation}.

\vspace{-5mm}
\paragraph{Minimax Regret}
The min-max regret as a learning objective has been considered in \parencite{agarwal2022minimax}.
In particular, based on a single training population $\bbP_{\train}$, they considered a pre-specified family of testing populations $\cP_{\test}$ to generalize to, and proposed to the min-max regret 
\begin{align}
	\min_{\theta\in\Theta}\sup_{\bbP_{\test}\in\cP_{\test}}R(\theta,\bbP_{\test}).
	\label{eq:MMR_AZ}
\end{align}
This is different from our motivation that given the training information of $\bbP^{(1)},\cdots,\bbP^{(K)}$, we aim to generalize to an unseen testing population $\bbP_{\test}$.

In terms of our ex-post generalization criterion \eqref{eq:regret_post}
where 
$\bbP_{\test}\in\{\bbP^{(1)},\cdots,\bbP^{(K)}\}$, 
our ex-post MMR reduces to the same mathematical formulation as \eqref{eq:MMR_AZ} by letting $\{\bbP^{(k)}\}_{k=1}^{K}=\cP_{\test}$. 
In terms of training, \parencite{agarwal2022minimax} proposed to solve the stochastic problem \eqref{eq:MMR_AZ} directly via on-demand querying from $\cP_{\test}$. This may not be suitable in our case, since our observed training samples $\cD^{(1)},\cdots,\cD^{(K)}$ are offline available.

In our ex-ante MMR \eqref{eq:regret_ante}, 
we have considered $\cP$ to incorporate the data distributions that are ex-ante realizable during training, but not every data distribution in $\cP$ is observed ex post.
Instead, we have the only access to the observed training data-distributions $\{\bbP^{(k)}\}_{k=1}^{K}$. This is different form the assumption in \parencite{agarwal2022minimax} that every $\bbP_{\test}\in\cP_{\test}$ can be realized from data. 
Therefore, our ex-ante MMR aims to solve a different problem compared to \parencite{agarwal2022minimax}.
}

\vspace{-7mm}
\subsection{Algorithm}\label{sec:algo}
\vspace{-2mm}
In this section, we consider the algorithm to solve the empirical MMR problem \eqref{eq:MMR}, 
where the empirical regret is denoted as $R_{k}(\theta) := R\big( \theta,\bbP_{n_{k}}^{(k)} \big)$ for ease of notation.
We assume that each $R_{k}(\cdot)$ is Lipschitz-gradient and strongly convex, which incorporates the linear regression and GLM settings in Sections \ref{sec:LM} and \ref{sec:GLM}.
For simplicity, consider $\Theta=\bbR^{p}$ and $\| \cdot \|_{2}$ as the $\ell^{2}$-norm on $\bbR^{p}$.
{\new
Following the linearization strategy in \parencite[Section 2.3]{nesterov2018lectures}, we iteratively solve
\begin{align}
	\theta^{(t)} \in \argmin_{\theta\in\bbR^{p}}\max_{\gamma\in\Delta^{K-1}}\sum_{k=1}^{K}\gamma_{k}\left\{ R_{k}\big( \theta^{(t-1)} \big) + \big\langle \nabla R_{k}\big( \theta^{(t-1)} \big), \theta - \theta^{(t-1)} \big\rangle + {L \over 2}\big\| \theta - \theta^{(t-1)} \big\|_{2}^{2} \right\}
	\label{eq:linearization}
\end{align}
for $t=1,2,\cdots$,
where $L > 0$ is the linearization constant.
Note that \eqref{eq:linearization} is a strongly-convex-concave (SC-C) bilinear game \parencite{chambolle2011first}, and can be equivalent to a quadratic programming (QP) with respect to $\gamma\in\Delta^{K-1}$.
The optimization is summarized in Algorithm \ref{algo:GD}.
In Supplementary Material D, we discuss more details on its motivations and the relationships with other optimization methods in the related literature.

\begin{algorithm}[!t]
	\SetKwInOut{Input}{Input}
	\SetKwInOut{Output}{Output}
	\caption{Linearization-Based Method for MMR}\label{algo:GD}
	\Input{Samples $\big\{Z_{i}^{(k)}\big\}_{i=1}^{n_{k}}$ for $k=1,\cdots,K$, 
		initialized estimate $\theta^{(0)} \in \bbR^{p}$, 
		linearization constant $L > 0$, total number of iterations $T$.}
	
	For $k=1,\cdots,K$, solve the within-sample ERM problem: 
	$\displaystyle\fR_{k}^{\dagger} = \min_{\beta \in \Theta}{1 \over n_{k}}\sum_{i=1}^{n_{k}}\ell_{\beta}\big( Z_{i}^{(k)} \big)$\;
	\For{$t = 1,\cdots,T$}{
		For $k=1,\cdots,K$, \\
		\, compute 
		$R_{k} = {1 \over n_{k}}\sum_{i=1}^{n_{k}}\ell_{\theta^{(t-1)}}(Z_{i}^{(k)}) - \fR_{k}^{\dagger}$ and $\nabla_{k}={1 \over n_{k}}\sum_{i=1}^{n_{k}}\nabla\ell_{\theta^{(t-1)}}\big( Z_{i}^{(k)} \big)$\;
		Let $q = (R_{1},\cdots,R_{K})^{\intercal} \in \bbR^{p}$ and 
		$\sfG = [\nabla_{1},\cdots,\nabla_{K}] \in \bbR^{p \times K}$\;
		Solve the following QP for $\gamma^{(t-1)}$:
		\[\max_{\gamma \in \Delta^{K-1}}\left\{ q^{\intercal}\gamma - {1 \over 2L}\gamma^{\intercal}\sfG^{\intercal}\sfG\gamma \right\};\]\\
		Update $\theta^{(t)} = \theta^{(t-1)} - L^{-1}\sfG\gamma^{(t-1)}$\;
	}
	Solve the QP at $t = T+1$ for $\gamma^{(T)}$\;
	\Output{The MMR estimator $\theta^{(T)}$, the dual weight $\gamma^{(T)}$.}
\end{algorithm}

The optimization guarantee of Algorithm \ref{algo:GD} is provided below.
Without loss of generality, we assume that the loss is twice-differentiable, so that gradient Lipschitzness and strong convexity are equivalent to the boundedness of the Hessian's eigenvalues.

\vspace{-2mm}
\begin{asm}[Gradient Lipschitzness and Strong Convexity]\label{asm:strong}
	Assume that the loss function $\ell_{\theta}(Z)$ is twice-differentiable in $\theta$,  
	and consider the empirical Hessian $\sfH_{k}(\theta) := {1 \over n_{k}}\sum_{i=1}^{n_{k}}\nabla^{2}\ell_{\theta}\big( Z_{i}^{(k)} \big)$.
	Further assume that for every compact set $\Theta\subseteq\bbR^{p}$,
	the eigenvalues of $\sfH_{k}(\theta)$ for $\theta\in\Theta$ and $k=1,\cdots,K$ are bounded from below $\ubar{\lambda}_{\Theta} > 0$ and from above $ \widebar{\lambda}_{\Theta} < +\infty $, respectively.
\end{asm}
\vspace{-5mm}
\begin{prop}\label{prop:GD}
    Consider the compact set   
	$\Theta = \{ \theta\in\bbR^{p}: \| \theta - \theta^{\star} \|_{2} \le \| \theta^{(0)} - \theta^{\star} \|_{2} \}$.
	Under Assumption \ref{asm:strong},
	Algorithm~\ref{algo:GD} with $L=\widebar{\lambda}_{\Theta}$ after $T$ iterations satisfies:
	\[ \begin{aligned}
		\left\|\theta^{(T)} - \theta^{\star}\right\|_{2}^{2} 
		\le \left( {\kappa - 1 \over \kappa + 1} \right)^{T}\left\|\theta^{(0)} - \theta^{\star}\right\|_{2}^{2}; \quad
		\cR_{\emp}\big(\theta^{(T)}\big) - \cR_{\emp}(\theta^{\star}) 
        \le {L+\mu \over 2}\left( {\kappa-1 \over \kappa+1} \right)^{T}\left\|\theta^{(0)} - \theta^{\star}\right\|_{2}^{2},
	\end{aligned} \]
	where $\theta^{\star}$ is the unique solution to $\min_{\theta \in \Theta}\cR_{\emp}(\theta)$,
	and $\kappa = \widebar{\lambda}_{\Theta}/\ubar{\lambda}_{\Theta}$.
\end{prop}
\vspace{-2mm}

We remark that the optimization guarantee in terms of the objective function $\cR_{\emp}$ decays exponentially in $T$, which is due to the Lipschitz gradient and strong convexity of $R_{k}(\cdot)$. 
For comparison, the objective decay of 
sub-gradient descent on $\cR_{\emp}$ is $\cO(T^{-1/2})$,
and the decay of gradient descent-ascent \parencite[Section 5.2]{bubeck2015convex} based on Lipschitz-gradient and convex $R_{k}(\cdot)$ is $\cO(T^{-1})$. 
Despite the advantage in iteration complexity,
our Algorithm \ref{algo:GD} needs to solve \eqref{eq:linearization} (a SC-C bilinear game or a QP) per iterate. 
Nevertheless, it remains superior when the cost of solving \eqref{eq:linearization} is reasonable.
More discussions are given in Supplementary Material D.
}

\vspace{-7mm}
\section{MMR for Linear Regression}\label{sec:LM}
\label{Sec: method}
\vspace{-2mm}
Our empirical MMR \eqref{eq:MMR} is formulated for a general learning task with a given loss function $\ell_{\theta}(Z)$ and a parameter space $\Theta$.
In this section, we specifically study linear regression to gain more insights on its structural properties and its distinctions from existing estimators,
including the \textit{maximin effect} \parencite{meinshausen2015maximin} as another robust estimator for the regression problem.

Consider the data $Z=(\bX,Y)$, where $\bX\in\bbR^{p}$ and $Y\in\bbR$ are the covariate vector and response variable, respectively.
The loss function for linear regression is the square loss $\ell_{\theta}(\bX,Y) = (Y - \bX^{\intercal}\theta)^{2}$ with the parameter of interest $\theta \in \bbR^{p}$, and the parameter space is $\Theta=\bbR^{p}$.
The corresponding risk function under a data distribution $\bbP$ is $R^{\dagger}(\theta,\bbP) = \bbE_{\bbP}(Y - \bX^{\intercal}\theta)^{2}$, which is also known as the \textit{mean square error (MSE)}.
For a general data-generating distribution $\bbP$, we do not assume a well-specified linear model $\bbE_{\bbP}(Y|\bX)=\bX^{\intercal}\beta$. 
Instead, we consider the linear regression coefficient as the risk minimizer $\beta(\bbP) \in \argmin_{\beta\in\bbR^{p}}R^{\dagger}(\beta,\bbP)$ whenever it exists. 
Let 
$\Sigma:=\bbE_{\bbP}(\bX\bX^{\intercal})$, $\mu:=\bbE_{\bbP}(\bX Y)$, 
$\beta(\bbP):=\Sigma^{-1}\mu$. 
When $\Sigma$ is positive definite,
$\beta(\bbP)$ is the unique risk minimizer, and the MSE risk function becomes\footnote{
	For a vector $u$ and a square matrix $\sfW$ with compatible dimensions, we denote $\| u \|_{\sfW}^{2} = u^{\intercal}\sfW u$.
}
\begin{align}
	R^{\dagger}(\theta,\bbP) = \underbrace{\|\theta - \beta(\bbP)\|_{\Sigma}^{2}}_{\text{regret}~ 
	R(\theta,\bbP)} + \underbrace{R^{\dagger}(\beta(\bbP),\bbP)}_{\text{WUV}},
	\label{eq:LM}
\end{align}
where we denote the \textit{within-population unexplained variance (WUV)} as $\sigma^{2} := \min_{\beta\in\bbR^{p}}R^{\dagger}(\beta,\bbP) = R^{\dagger}(\beta(\bbP),\bbP)$.
We write $\beta=\beta(\bbP)$ when there is no ambiguity.
The decomposition in (\ref{eq:LM}) suggests that the risk function $R^{\dagger}(\theta,\bbP)$ depends on $\bbP$ through the population characteristics $(\beta,\Sigma,\sigma^{2})$, while the regret function $R(\theta,\bbP)$ depends on $\bbP$ through $(\beta,\Sigma)$ only.

During the training stage, we observe the samples $\cD^{(k)}=\{ \bX_{i}^{(k)}, Y_{i}^{(k)} \}_{i=1}^{n_{k}}$ from the training populations $\bbP^{(k)}$ for $k=1,\cdots,K$.
We denote $(\beta_{k},\Sigma_{k},\sigma_{k}^{2})$ as the training population characteristics of $\bbP^{(k)}$ as above, and the empirical characteristics $(\widehat{\beta}_{k},\widehat{\Sigma}_{k},\widehat{\sigma}_{k}^{2})$ as the corresponding empirical averages over $\cD^{(k)}$.
In particular, $\widehat{\beta}_{k}$ is the least-squares estimate on $\cD^{(k)}$.
Then the ex-post MMR based on \eqref{eq:regret_post} and empirical MMR based on \eqref{eq:regret_emp} are equivalent to
\begin{align}
	\min_{\theta\in\bbR^{p}}\left\{ \cR_{\post}(\theta) = \max_{1 \le k \le K}\| \theta - \beta_{k} \|_{\Sigma_{k}}^{2} \right\}; \quad
	\min_{\theta\in\bbR^{p}}\left\{ \cR_{\emp}(\theta) = \max_{1 \le k \le K}\| \theta - \widehat{\beta}_{k} \|_{\widehat{\Sigma}_{k}}^{2} \right\}.
	\label{eq:MMR_LM}
\end{align}
In particular, the ex-post MMR solves a min-max-distance problem,
and the empirical MMR can be considered as a plug-in analog of the ex-post MMR, where the population characteristics $(\beta_{k},\Sigma_{k})$ are substituted by the empirical estimates $(\widehat{\beta}_{k},\widehat{\Sigma}_{k})$.%

{\new
\vspace{-7mm}
\subsection{Comparisons of Robust Methods}
\label{sec:compare_LM}
\vspace{-2mm}
In this section, we discuss the relationships of several robust methods for linear regression with multiple heterogeneous populations. 
Based on the MSE risk, the GDRO problem \eqref{eq:GDRO} in Section~\ref{sec:compare} is equivalent to 
\begin{align}
	\min_{\theta\in\bbR^{p}}\max_{1 \le k \le K}\Big\{ \underbrace{\| \theta - \beta_{k} \|_{\Sigma_{k}}^{2}}_{\text{regret}} + \underbrace{\sigma_{k}^{2}}_{\text{WUV}} \Big\}.
	\label{eq:GDRO_LM}
\end{align}
Such a GDRO problem can be sensitive to the heterogeneous WUVs $\{\sigma_{k}^{2}\}_{k=1}^{K}$.
To mitigate this challenge, \textcite{meinshausen2015maximin} proposed 
the \textit{maximin explained variance (MMV)}:%
\footnote{%
	We have assumed $\bbE_{\bbP^{(k)}}(Y) = 0$, and $\bbE_{\bbP^{(k)}}(Y^{2})$ is the variance of $Y$.
}
\begin{align}
	\max_{\theta \in \bbR^{p}}\min_{1 \le k \le K}\Big\{ V(\theta,\bbP^{(k)})  := \bbE_{\bbP^{(k)}}[Y^{2} - (Y - \bX^{\intercal}\theta)^{2}] \Big\}.
	\label{eq:MMV}
\end{align}
Here, $V(\theta,\bbP)$ is the explained variance criterion as the MSE-difference $R^{\dagger}(0,\bbP) - R^{\dagger}(\theta,\bbP)$, where $\theta=0$ serves as a null reference.
It can be equivalently written as
\begin{align}
	\min_{\theta \in \bbR^{p}}\max_{1 \le k \le K}\Big\{ - V(\theta,\bbP^{(k)}) = \underbrace{\| \theta - \beta_{k} \|_{\Sigma_{k}}^{2}}_{\text{regret}} - \underbrace{\nu_{k}^{2}}_{\text{WEV}} \Big\},
	\label{eq:MMV_LM}
\end{align}
where we denote the \textit{within-population explained variance (WEV)}
as $\nu_{k}^{2} := \max_{\beta\in\bbR^{p}}V(\beta,\bbP^{(k)}) = V(\beta_{k},\bbP^{(k)}) = \| \beta_{k} \|_{\Sigma_{k}}^{2}$ for $k=1,\cdots,K$.
In particular, for the negative explained variance $-V(\theta,\bbP^{(k)})$ as a risk function of $\theta$ on $\bbP^{(k)}$, 
the corresponding WMR is $-\nu_{k}^{2}$.
This suggests that the MMV \eqref{eq:MMV_LM} avoids the dependency on the WUVs $\{\sigma_{k}^{2}\}_{k=1}^{K}$,
but alternatively involves the negative WEVs $\{-\nu_{k}^{2}\}_{k=1}^{K}$ as the within population minimized risks.

\begin{table}[!ht]
	\centering
	\caption{Comparisons of GDRO, MMV and MMR in Linear Regression}\label{tab:criteria}
	\resizebox{.9\textwidth}{!}{%
		\begin{tabular}{c||c|c}
		\hline\hline
		\textbf{Method}
		& \makecell{\textbf{Within-Population} \\ \textbf{Minimized Risk (WMR)}}
		& \makecell{\textbf{Characterization of Degeneration} \\ $\theta^{\star} = \beta_{k^{\star}}$ for some $k^{\star}$; 
		$\Delta_{k,k^{\star}} := \| \beta_{k^{\star}} - \beta_{k} \|_{\Sigma_{k}}^{2}$} \\
		\hline
		GDRO \eqref{eq:GDRO_LM}
		& unexplained variance $\sigma_{k}^{2}$
		& \makecell{
			$\displaystyle\sigma_{k^{\star}}^{2} \ge \max_{k\neq k^{\star}}\{\sigma_{k}^{2} + \Delta_{k,k^{\star}}\}$ 
		} \\
		\hline
		MMV \eqref{eq:MMV_LM}
		& negative explained variance $-\nu_{k}^{2}$
		& \makecell{
			$\displaystyle\nu_{k^{\star}}^{2} \le \min_{k\neq k^{\star}}\{\nu_{k}^{2} - \Delta_{k,k^{\star}}\}$ 
		} \\
		\hline
		\textbf{MMR} \eqref{eq:MMR_LM}
		& $0$
		& \makecell{
			$\beta_{1}=\cdots=\beta_{K}$ 
			\textbf{\textit{(homogeneity)}}
		} \\
		\hline\hline
	\end{tabular}
	}
    \begin{minipage}{.9\textwidth}
        \scriptsize\it
        \underline{Note.} The characterizations of degeneration are based on Propositions C.1 and C.2 in Supplementary Material C.
    \end{minipage}
\end{table}

The GDRO \eqref{eq:GDRO_LM}, MMV \eqref{eq:MMV_LM} and MMR \eqref{eq:MMR_LM} can be unified as the min-max of (regret + WMR), 
which is summarized in Table \ref{tab:criteria}.
These methods can be considered as instances of the general GDRO \eqref{eq:GDRO} under different risk functions,
and hence could be sensitive to the heterogeneous WMRs.
In the extreme scenario, the min-max estimator $\theta^{\star}$ can reduce to the regression coefficient $\beta_{k^{\star}}$ of a single dominating population $\bbP^{(k^{\star})}$, which we refer to as \textbf{degeneration}.
In particular, a degenerate min-max estimator $\theta^{\star}=\beta_{k^{\star}}$ can be conservative or uninformative for its performance on the remaining training populations $\{\bbP^{(k)}\}_{k\neq k^{\star}}$.
As in Table \ref{tab:criteria},
the GDRO degenerates whenever some WUV $\sigma_{k^{\star}}^{2}$ is sufficiently large,
while the MMV degenerates whenever some WEV $\nu_{k^{\star}}^{2}$ is sufficiently small.
Both degeneration corresponds to the scenario that, even not to consider the generalization to the other populations, the best linear regression fit on the dominating population $\bbP^{(k^{\star})}$ is much poorer than the model fits on the other populations.
As an example, if the conditional variance of $Y|\bX$ on $\bbP^{(k^{\star})}$ is much larger than that on $\{\bbP^{(k)}\}_{k\neq k^{\star}}$, then $\bbP^{(k^{\star})}$ can dominate the GDRO.
As another example, if the linear relationship between $Y$ and $\bX$ on $\bbP^{(k^{\star})}$ is much weaker than that on $\{\bbP^{(k)}\}_{k\neq k^{\star}}$, then $\bbP^{(k^{\star})}$ can dominate the MMV.

The MMR has zero WMRs across all training populations,
and hence is insensitive to the heterogeneous WUVs $\{\sigma_{k}^{2}\}_{k=1}^{K}$ and WEVs $\{\nu_{k}^{2}\}_{k=1}^{K}$.
Moreover, the degeneration happens only when the training regression coefficients are \textbf{homogeneous},
in which case the MMR estimator $\theta^{\star}$ is a simultaneous risk minimizer on all training populations $\bbP^{(1)},\cdots,\bbP^{(K)}$.
Whenever the heterogeneity exists, $\theta^{\star}$ would not be dominated by a single training population. 

In Section \ref{subsec: simu lm}, we provide numerical comparisons of GDRO, MMV and MMR for their sensitivity to the heterogeneous WUVs and WEVs.
In Supplementary Material E.1, we provide additional discussions on the transformation-equivariance enjoyed by the GDRO and MMR but not by MMV.
This property helps the interpretation of the resulting estimator as a common effect parameter across heterogeneous populations.
}

\vspace{-7mm}
\subsection{Geometric Characterization of MMR and MMV}
\label{sec:MMV_MMR}
\begin{figure}[!t]
	\centering
	\includegraphics[width=0.4\textwidth]{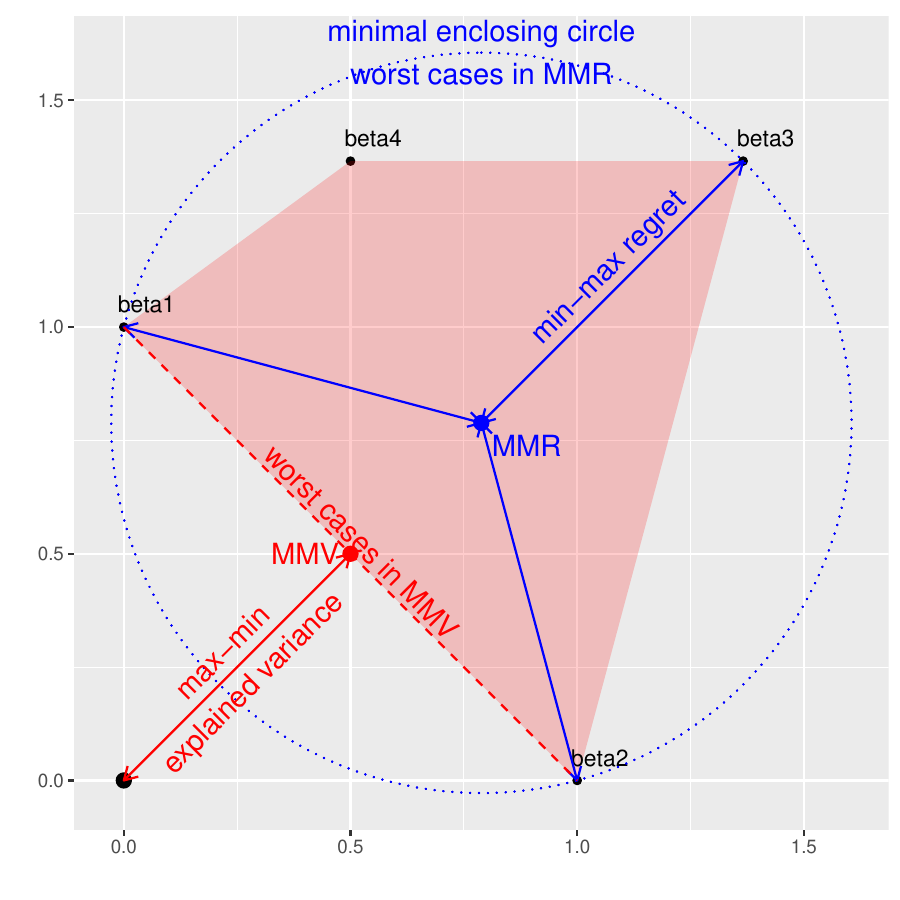}
	\includegraphics[width=0.4\textwidth]{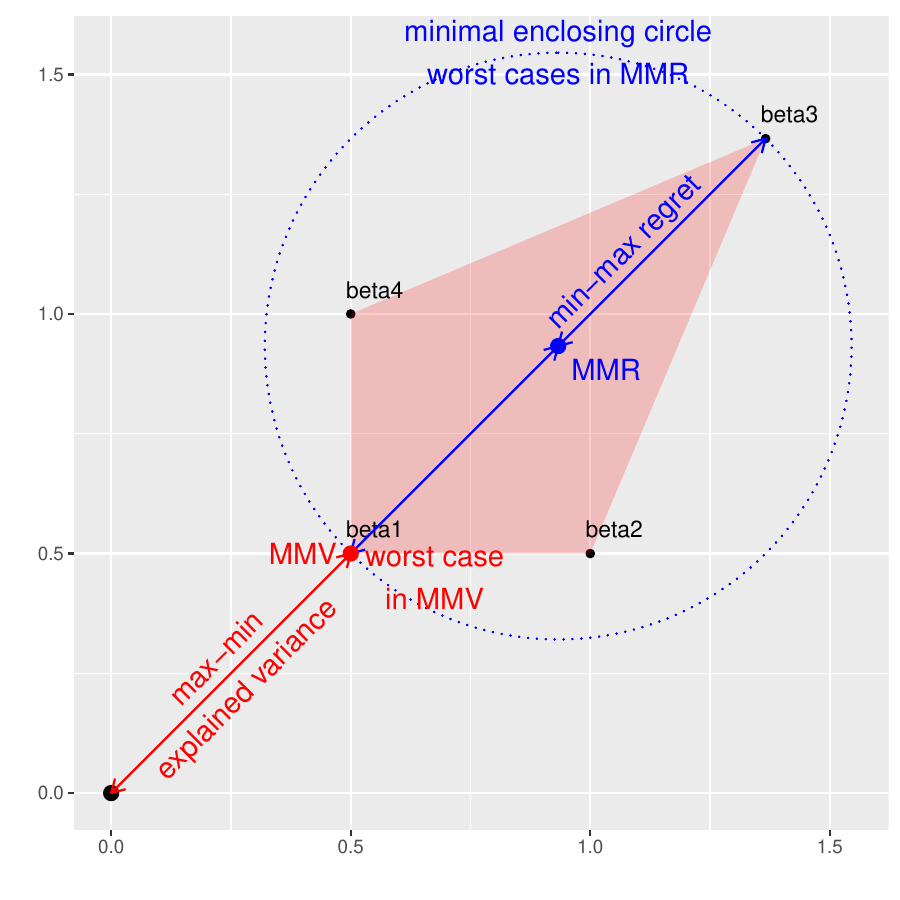}
	\vspace{-2mm}
	\singlespacing
	\caption{Illustration of MMR and MMV for linear regression based on four population regression coefficients $\beta_{1},\beta_{2},\beta_{3},\beta_{4} \in \bbR^{2}$ and the identity covariate covariance matrix $\Sigma$. 
    The MMV estimator is the closest point from $\cB = \conv\{\beta_{1},\beta_{2},\beta_{3},\beta_{4}\}$ to the origin.
    The MMR estimator is the centroid of the minimal enclosing circle of $\cB$. 
    The squared distance between $\cB$ and the origin is the max-min explained variance.
    The squared radius of the enclosing circle is the min-max regret.
    In the left plot, the MMV supporting coefficients are $\beta_{1},\beta_{2}$,
    and the MMR supporting coefficients are $\beta_{1},\beta_{2},\beta_{3}$.
    In the right plot, the MMV estimator degenerates to $\beta_{1}$.
    The MMR does not degenerate, with $\beta_{1},\beta_{2}$ as the supporting coefficients.
	}
	\vspace{-4mm}
	\label{fig:MMV_MMR}
\end{figure}

\vspace{-2mm}
In this section, we further compare the MMR \eqref{eq:MMR_LM} and MMV \eqref{eq:MMV_LM} via geometric characterization. 
To facilitate our discussion, 
we assume that the training populations $\bbP^{(1)},\cdots,\bbP^{(K)}$ share a common covariate covariance matrix $\Sigma_{1}=\cdots=\Sigma_{K}\equiv \Sigma$, which is positive definite.
The training populations are characterized by their regression coefficients $\beta_{1},\cdots,\beta_{K}$.
We illustrate the comparison in two toy examples given in Figure \ref{fig:MMV_MMR}. 
It suggests the distinctions between MMV and MMR in terms of their estimators and the supporting coefficients,
where the estimators are determined by the supporting coefficients via convex aggregation.
These structural results are investigated in Theorems \ref{thm:MMR_LM}-\ref{thm:MMV} via duality in this section.

We first characterize the ex-post MMR problem \eqref{eq:MMR_LM} for linear regression.
As a min-max-distance problem, it can be equivalently written as
\begin{align}
	\min_{\theta \in \bbR^{p}, R \ge 0}\Big\{ R ~ \text{subject to} ~ \|\theta - \beta_{k}\|_{\Sigma}^{2} \le R, ~ 1 \le k \le K \Big\}.
	\label{eq:MMR_LM_ball}
\end{align}
Problem \eqref{eq:MMR_LM_ball} aims to find the \textbf{minimal ellipsoid} $\cE_{\Sigma}(\theta^{\star},R^{\star}) = \{ \beta\in\bbR^{p}: \| \beta-\theta^{\star} \|_{\Sigma}^{2} \le R^{\star} \}$ \textbf{enclosing} $\{\beta_{k}\}_{k=1}^{K}$\footnote{
    Note that $\cE_{\Sigma}(\theta^{\star},R^{\star})$ is a convex set in $\bbR^{p}$. 
    It is also the minimal ellipsoid enclosing $\conv\{\beta_{k}\}_{k=1}^{K}$.
}, 
with 
$\theta^{\star}$ and $R^{\star}$ as the optimized centroid and squared radius, respectively.
Denote $\partial\cE_{\Sigma}(\theta^{\star},R^{\star}) = \{ \beta\in\bbR^{p}: \| \beta-\theta^{\star} \|_{\Sigma}^{2} = R^{\star} \}$ as its boundary. 
The MMR solution pair $(\theta^{\star},R^{\star})$ is further characterized below.

\vspace{-2mm}
\begin{thm}[Characterization of MMR]\label{thm:MMR_LM}
	Consider the training population characteristics $(\beta_{k},\Sigma_{k})$ for $k=1,\cdots,K$ in linear regression.
	Assume that $\Sigma_{1} = \cdots = \Sigma_{K} \equiv \Sigma$, which is positive definite. 
	Then the ex-post MMR solution to \eqref{eq:MMR_LM} is 
	$\theta^{\star} = \sum_{k=1}^{K}\gamma_{k}^{\star}\beta_{k}$,
	where 
	\begin{align}
		\gamma^{\star}\in 
		\argmax_{\gamma\in\Delta^{K-1}}\left\{
		\fR(\gamma)
		= \sum_{k=1}^{K}\gamma_{k}\| \beta_{k} \|_{\Sigma}^{2} - \left\| \sum_{k=1}^{K}\gamma_{k}\beta_{k} \right\|_{\Sigma}^{2} \right\}. 
		\label{eq:MMR_LM_dual}
	\end{align}
	The min-max regret is $R^{\star} = \min_{\theta\in\bbR^{p}}\max_{1 \le k \le K}R(\theta,\bbP^{(k)}) = \fR(\gamma^{\star})$, 
	and we have $\{\beta_{k}\}_{k=1}^{K} \subseteq \cE_{\Sigma}(\theta^{\star},R^{\star})$.
	For any data-distribution $\bbQ$ with $\bbE_{\bbQ}(\bX\bX^{\intercal}) = \Sigma$ 
	and $\beta(\bbQ) = \Sigma^{-1}\bbE_{\bbQ}(\bX Y)\in\cE_{\Sigma}(\theta^{\star},R^{\star})$, 
	we have 
	$R(\theta^{\star},\bbQ)\le R^{\star}$, with equality if and only if $\beta(\bbQ)\in\partial\cE_{\Sigma}(\theta^{\star},R^{\star})$.
    Moreover, 
    the supporting set satisfies $\cK^{\star}
    =\{k:\beta_{k}\in\partial\cE_{\Sigma}(\theta^{\star},R^{\star})\} \supseteq \{k:\gamma_{k}^{\star} > 0\}$.
\end{thm}
\vspace{-2mm}

Theorem \ref{thm:MMR_LM} suggests that the MMR estimator $\theta^{\star}$ is a convex aggregation of the training regression coefficients $\{\beta_{k}\}_{k=1}^{K}$, and the aggregation weight $\gamma^{\star}$ is optimized from the \textbf{dual MMR problem} \eqref{eq:MMR_LM_dual}.
In particular, the dual MMR problem solves a \textbf{robust ellipsoid} $\cE_{\Sigma}(\theta^{\star},R^{\star})$,
such that for any data-distribution $\bbQ$ with regression coefficient $\beta(\bbQ)\in\cE_{\Sigma}(\theta^{\star},R^{\star})$, 
the MMR estimator $\theta^{\star}$ incurs a regret $R(\theta^{\star},\bbQ)$ upper bounded by $R^{\star}$.
This includes all training populations $\{\bbP^{(k)}\}_{k=1}^{K}$ whose regression coefficients $\{\beta_{k}\}_{k=1}^{K}$ are contained in $\cE_{\Sigma}(\theta^{\star},R^{\star})$.
The ellipsoid boundary $\partial\cE_{\Sigma}(\theta^{\star},R^{\star})$ further contains the \textbf{supporting coefficients} $\{\beta_{k}\}_{k\in\cK^{\star}}$, whose convex aggregation determines the MMR estimator $\theta^{\star} = \sum_{k\in\cK^{\star}}\gamma_{k}^{\star}\beta_{k}$.
They correspond to the worst-case training populations in MMR, such that the regrets of $\theta^{\star}$ on these populations $R(\theta^{\star},\bbP^{(k)})$ for $k\in\cK^{\star}$ attain the worst-case upper bound $R^{\star}$.

For comparison, we also characterize the MMV problem \eqref{eq:MMV_LM}.
For $a\in\bbR^{p}$ and $b\in\bbR$,
we denote an upper half-space as $\cH_{\Sigma}^{\ge}(a,b) = \{u\in\bbR^{p}: a^{\intercal}\Sigma u \ge b \}$, 
and its boundary hyperplane as $\cH_{\Sigma}^{=}(a,b) = \{u\in\bbR^{p}: a^{\intercal}\Sigma u=b\}$.

\vspace{-2mm}
\begin{thm}[Characterization of MMV\footnote{
    The dual MMV problem \eqref{eq:MMV_dual} was also characterized in the prior literature \parencite{meinshausen2015maximin,guo2023statistical}, where \eqref{eq:MMV_dual} can be equivalently written as $\min\big\{\|\theta\|_{\Sigma}^{2}: \theta \in \conv\{\beta_{k}\}_{k=1}^{K} \big\}$.
    Our Theorem \ref{thm:MMV} further characterizes the robust half-space $\cH_{\Sigma}^{\ge}(\theta^{\star},V^{\star})$ and its boundary hyperplane $\cH_{\Sigma}^{=}(\theta^{\star},V^{\star})$ that contains the supporting coefficients.
}]\label{thm:MMV}
	Consider the training population characteristics $(\beta_{k},\Sigma_{k})$ for $k=1,\cdots,K$ in linear regression.
	Assume that $\Sigma_{1} = \cdots = \Sigma_{K} \equiv \Sigma$, which is positive definite. 
	Then the MMV solution to \eqref{eq:MMV} is 
	$\theta^{\star} = \sum_{k=1}^{K}\gamma_{k}^{\star}\beta_{k}$,
	where 
	\begin{align}
		\gamma^{\star}\in 
		\argmin_{\gamma\in\Delta^{K-1}}\left\{ \fV(\gamma) := \left\| \sum_{k=1}^{K}\gamma_{k}\beta_{k} \right\|_{\Sigma}^{2} \right\}.
		\label{eq:MMV_dual}
	\end{align}
	The max-min explained variance is 
    $V^{\star} = \max_{\theta\in\bbR^{p}}\min_{1 \le k \le K}V(\theta,\bbP^{(k)}) = \fV(\gamma^{\star})$, 
	and we have $\{\beta_{k}\}_{k=1}^{K} \subseteq \cH_{\Sigma}^{\ge}(\theta^{\star},V^{\star})$.
	For any data-distribution $\bbQ$ with $\bbE_{\bbQ}(\bX\bX^{\intercal}) = \Sigma$ 
	and $\beta(\bbQ) = \Sigma^{-1}\bbE_{\bbQ}(\bX Y)\in\cH_{\Sigma}^{\ge}(\theta^{\star},V^{\star})$, 
	we have 
	$V(\theta^{\star},\bbQ)\ge V^{\star}$, with equality if and only if $\beta(\bbQ)\in\cH_{\Sigma}^{=}(\theta^{\star},V^{\star})$.
    The supporting set satisfies 
    $\cK^{\star}
    =\{k:\beta_{k}\in\cH_{\Sigma}^{=}(\theta^{\star},V^{\star})\} \supseteq \{k:\gamma_{k}^{\star} > 0\}$.
\end{thm}
\vspace{-2mm}

Comparing Theorems \ref{thm:MMR_LM} and \ref{thm:MMV}, 
the main distinction of MMV is that the \textbf{dual MMV problem} \eqref{eq:MMV_dual} solves a \textbf{robust half-space} $\cH_{\Sigma}^{\ge}(\theta^{\star},V^{\star})$. 
Such a dual problem is equivalent to minimizing the WEV among the mixtures of $\{\bbP^{(k)}\}_{k=1}^{K}$. 
Specifically, 
we denote $\nu^{2}(\bbQ) := \max_{\beta\in\bbR^{p}}V(\beta,\bbQ)$ as the WEV functional of the data-distribution $\bbQ$,
and in particular, $\nu^{2}(\bbP^{(k)}) = \nu_{k}^{2}$ for $k=1,\cdots,K$.
Then \eqref{eq:MMV_dual} is equivalent to minimizing $\nu^{2}(\bbQ)$ over $\bbQ=\sum_{k=1}^{K}\gamma_{k}\bbP^{(k)}$, $\gamma\in\Delta^{K-1}$.\footnote{
    This is also a consequence of our characterization of the general dual GDRO problem in Supplementary Material C. 
}
As a consequence, MMV can be sensitive to the heterogeneity of the training WEVs $\{\nu_{k}^{2}\}_{k=1}^{K}$.
In particular, $V^{\star} \le \min_{1 \le k \le K}\nu_{k}^{2}$, and a sufficiently small $\nu_{k^{\star}}^{2}$ could dominate the dual MMV, 
leading to the degeneration in Section \ref{sec:compare_LM}.

{\new
\vspace{-7mm}
\section{MMR for Generalized Linear Model (GLM)}\label{sec:GLM}
\vspace{-2mm}
In this section, we extend Section \ref{sec:LM} to a broader class of applications,
where the loss function is motivated from the generalized linear model (GLM) with a canonical link \parencite{mccullagh2019generalized}. 
Specifically, 
consider the data $Z=(\bX,Y)\in\bbR^{p}\times\bbR$,
the parameter space $\Theta=\bbR^{p}$,
and a three-times differentiable strictly convex function $\cA: \bbR \to \bbR$.
The GLM-likelihood-based loss and risk functions are:\footnote{
    It is motivated from the GLM density in the canonical form: $p_{\theta}(y|\bx) = h(y,\sigma)\exp\big\{ {(y\bx^{\intercal}\theta - \cA(\bx^{\intercal}\theta)) / \sigma} \big\}$,
    where 
    $\theta \in \bbR^{p}$ is the parameter of interest, 
    $\sigma > 0$ is the dispersion parameter, 
    $h(y,\sigma)$ does not depend on $\theta$.
    The log-likelihood is
    $\log p_{\theta}(y|\bx) = \{y\bx^{\intercal}\theta - \cA(\bx^{\intercal}\theta)\}/\sigma + \log h(y,\sigma)$.
    In particular, \eqref{eq:GLM} is the negative log-likelihood without $\sigma$ and $\log h(y,\sigma)$.
	An alternative definition of the loss and risk functions under GLM can be based on the deviance \parencite{mccullagh2019generalized}. 
	More discussions are provided in Supplementary Material F.2.
}
\begin{align}
	\begin{aligned}
		& \ell_{\theta}(\bx,y)
		= \cA(\bx^{\intercal}\theta) - y\bx^{\intercal}\theta; \quad
		R^{\dagger}(\theta,\bbP) 
		= \sfA(\theta) - \langle \mu,\theta \rangle, \\
		& \text{where} \quad 
		\sfA(\theta) 
		:= \bbE_{\bbP}[\cA(\bX^{\intercal}\theta)]; \quad
		\mu 
		:= \bbE_{\bbP}(\bX Y).
	\end{aligned}
	\label{eq:GLM}
\end{align}
For linear regression, $\cA(\eta) = \eta^{2}/2$
and $\sfA(\theta) = \theta^{\intercal}\Sigma\theta/2$ for $\Sigma = \bbE_{\bbP}(\bX\bX^{\intercal})$.
The likelihood-based risk function becomes $R^{\dagger}(\theta,\bbP) = -(1/2)V(\theta,\bbP)$, where $V(\theta,\bbP)$ is the explained variance in \eqref{eq:MMV}.
For logistic regression, $\cA(\eta)=\log(1+e^{\eta})$. 
More examples are provided in Supplementary Material F.1.
To extend Section \ref{sec:LM}, we first discuss the GLM risk minimization on a single population, and characterize the regret function in Section \ref{sec:GLM_regret}. Then we discuss the MMR problem in Section \ref{sec:GLM_MMR}, and its geometric characterization in Section \ref{sec:GLM_geo}.

\vspace{-7mm}
\subsection{Regret on a Single Population}\label{sec:GLM_regret}
\vspace{-2mm}
For a general data-generating distribution $\bbP$, we do not assume that the density of $Y|\bX$ is a well-specified GLM. Instead, we consider the GLM parameter as the risk minimizer $\beta(\bbP) \in \argmin_{\beta\in\bbR^{p}}R^{\dagger}(\beta,\bbP)$ whenever it exists.
It corresponds to the GLM with the minimal Kullback-Leibler divergence relative to the data-distribution $\bbP$ \parencite{white1982maximum}.
To ensure the existence and uniqueness of $\beta(\bbP)$,
we assume the regularity conditions for $\bbP$ as in \parencite{white1982maximum}.

\vspace{-2mm}
\begin{asm}[Regularity of Data-Distribution]\label{asm:reg_GLM}
	(a) $\Sigma:=\bbE_{\bbP}(\bX\bX^{\intercal})$ is positive definite. 
	(b) 
	$\theta\mapsto R^{\dagger}(\theta,\bbP)$ has compact sub-level sets.\footnote{
        For every $r \in \bbR$, $\{\theta\in\bbR^{p}: R^{\dagger}(\theta,\bbP) \le r \}$ is a sub-level set of $R^{\dagger}(\cdot,\bbP)$.
        The compactness of the risk sub-level sets is equivalent to $\lim_{\|\theta\|_{2}\to+\infty}R^{\dagger}(\theta,\bbP) = +\infty$.
		In logistic regression, such a compactness rules out the scenario that $\bX|(Y=1)$ and $\bX|(Y=0)$ are linearly separable under $\bbP$ \parencite{ji2019implicit}, 
		in which case the GLM risk infimum $\inf_{\beta\in\bbR^{p}}R^{\dagger}(\beta,\bbP) = 0$ is not attainable and can be approached as $\|\beta\|_{2} \to +\infty$.
	}
	(c) 
	For every compact set $\Theta \subseteq \bbR^{p}$, 
	we have uniformly for $\theta\in\Theta$,
	$|\cA(\bX^{\intercal}\theta)|$,
	$\| \bX \|_{2} |Y|$,
	$\| \bX \|_{2} |\cA'(\bX^{\intercal}\theta)|$,
	$\| \bX \|_{2}^{2} \cA''(\bX^{\intercal}\theta)$,
	$\| \bX \|_{2}^{3} |\cA'''(\bX^{\intercal}\theta)|$
	are upper bounded by some $B_{\Theta}(\bX,Y)$ such that $\bbE_{\bbP}[B_{\Theta}(\bX,Y)] < +\infty$.
\end{asm}
\vspace{-2mm}

Define the population Hessian 
$\cI(\theta) := \nabla_{\theta\theta^{\intercal}}^{2}R^{\dagger}(\theta,\bbP) 
= \nabla^{2}\sfA(\theta)$.
It is also the Fisher information of a well-specified GLM with parameter $\theta$.
In the following Proposition \ref{prop:GLM}, 
we establish its positive definiteness, which implies the existence and uniqueness of the GLM risk minimizer.
Based on $\sfA(\cdot)$ as the distance-generating function on $\bbR^{p}$,
we further consider the Bregman divergence
$\sfD_{\sfA}(\theta_{0}\|\theta_{1}) := \sfA(\theta_{1}) - \sfA(\theta_{0}) - \langle \nabla\sfA(\theta_{0}), \theta_{1} - \theta_{0} \rangle$
for $\theta_{0},\theta_{1} \in \bbR^{p}$, which is an asymmetric distance of $\theta_{1}$ relative to $\theta_{0}$. 
Based on the convex conjugate 
$\sfA^{*}(\theta^{*}) := \sup_{\theta\in\bbR^{p}}\{\langle \theta^{*},\theta \rangle - \sfA(\theta)\}$ as a distance-generating function of $\theta^{*}\in\bbR^{p}$, we also have 
$\sfD_{\sfA}(\theta_{0}\|\theta_{1})=\sfD_{\sfA^{*}}(\theta_{1}^{*}\|\theta_{0}^{*})$, where $\theta_{j}^{*} = \nabla\sfA(\theta_{j})$ $\Leftrightarrow$ $\theta_{j} = \nabla\sfA^{*}(\theta_{j}^{*})$ for $j=0,1$.
These are related to the GLM regret in Proposition \ref{prop:GLM} below.

\vspace{-2mm}
\begin{prop}[GLM under General Specification]\label{prop:GLM}
	Consider the GLM risk $R^{\dagger}(\theta,\bbP)$ in \eqref{eq:GLM}.
	Fix a data-generating distribution $\bbP$ satisfying Assumption \ref{asm:reg_GLM}.
	We have the followings.
    \begin{enumerate}[label=(\alph*),leftmargin=*,itemsep=0mm]
        \vspace{-2mm}
		\item\label{item:GLM_mono} The Hessian $\cI(\theta) = \bbE_{\bbP}[\bX\bX^{\intercal}\cA''(\bX^{\intercal}\theta)]$ is positive definite and continuous in $\theta$.
		\item\label{item:GLM_id} $\inf_{\beta\in\bbR^{p}}R^{\dagger}(\beta,\bbP)$ is attained at $\beta=\beta(\bbP)$, which is the unique solution to
		$\bbE_{\bbP}\big\{ \bX[Y-\cA'(\bX^{\intercal}\beta)] \big\} = 0$,
		and equivalently, 
		$\mu = \nabla\sfA(\beta)$ $\Leftrightarrow$ $\beta = \nabla\sfA^{*}(\mu)$.
		\item The regret function
		is 
		$R(\theta,\bbP) = \sfD_{\sfA}(\beta\|\theta) = \sfD_{\sfA^{*}}(\theta^{*}\|\mu)$ for any $\theta\in\bbR^{p}$ and $\theta^{*} = \nabla\sfA(\theta)$.
	\end{enumerate}
\end{prop}
\vspace{-2mm}

Proposition \ref{prop:GLM} establishes the identification of GLM parameter under general specification, that is, $\beta = \nabla\sfA^{*}(\mu)$. 
If the GLM is well-specified, then
$\cA'(\bX^{\intercal}\beta)=\bbE_{\bbP}(Y|\bX)$.
When the GLM is mis-specified,
$\cA'(\bX^{\intercal}\beta)$ may not be the same as $\bbE_{\bbP}(Y|\bX)$, but can still be interpreted as a GLM-based prediction of $Y$.
Proposition \ref{prop:GLM} also establishes that the GLM regret of $\theta$ is a Bregman divergence of $\theta$ relative to the GLM parameter $\beta$. 
Such a divergence is equivalent to the squared distance $\|\theta -\beta\|_{2}^{2}$ up to some factors for $\theta$ in a compact set. See Supplementary Material Lemma F.1.
In the linear regression case, we specifically have $R(\theta,\bbP) = \sfD_{\sfA}(\beta\|\theta) = (1/2)\| \theta - \beta \|_{\Sigma}^{2}$ as in \eqref{eq:LM}.

\vspace{-7mm}
\subsection{MMR across Multiple Populations}\label{sec:GLM_MMR}
\vspace{-2mm}
Consider $\cD^{(k)} = \{\bX_{i}^{(k)}, Y_{i}^{(k)}\}_{k=1}^{K}$ for $k=1,\cdots,K$, and the corresponding training populations $\bbP^{(1)},\cdots,\bbP^{(K)}$.
For the $k$-th data distribution $\bbP^{(k)}$ satisfying Assumption \ref{asm:reg_GLM},
we denote the population characteristics 
$(\beta_{k},\mu_{k},\sfA_{k})$,
where 
$\mu_{k} = \bbE_{\bbP^{(k)}}(\bX Y)$,
$\sfA_{k}(\theta) = \bbE_{\bbP^{(k)}}[\cA(\bX^{\intercal}\theta)]$,
and $\beta_{k} = \nabla\sfA_{k}^{*}(\mu_{k})$ as in Proposition \ref{prop:GLM}.
We refer to $\beta_{1},\cdots,\beta_{K}$ as the local GLM parameters.
Based on the $k$-th sample $\cD^{(k)}$, 
the respective empirical characteristics 
$(\widehat{\beta}_{k},\widehat{\mu}_{k},\widehat{\sfA}_{k})$ 
are the empirical averages over $\cD^{(k)}$.
In particular, $\widehat{\beta}_{k}$ is the maximum likelihood estimate on $\cD^{(k)}$.

\vspace{-2mm}
\begin{coro}[GLM-Based MMR]\label{coro:MMR_GLM}
	Suppose that for $k=1,\cdots,K$, $\bbP^{(k)}$ satisfies Assumption~\ref{asm:reg_GLM}.
	Then the GLM-based ex-post MMR problem \eqref{eq:regret_post} is equivalent to
	$\displaystyle
    \min_{\theta\in\bbR^{p}}\max_{1 \le k \le K}\sfD_{\sfA_{k}}(\beta_{k}\|\theta)$.
	The GLM-based empirical MMR problem \eqref{eq:MMR} is equivalent to
	$\displaystyle
    \min_{\theta\in\bbR^{p}}\max_{1 \le k \le K}\sfD_{\widehat{\sfA}_{k}}(\widehat{\beta}_{k}\|\theta)$.
\end{coro}
\vspace{-2mm}

Corollary \ref{coro:MMR_GLM} suggests that the ex-post MMR solves a min-max-distance problem,
and the empirical MMR \eqref{eq:MMR} can be considered as a plug-in analog of the ex-post MMR, where the population characteristics $(\beta_{k},\sfA_{k})$ are substituted by the empirical estimates $(\widehat{\beta}_{k},\widehat{\sfA}_{k})$.
In linear regression, the ex-post and empirical MMR problems are equivalent to \eqref{eq:MMR_LM},
and the GDRO \eqref{eq:GDRO} based on the GLM-likelihood-based risk in \eqref{eq:GLM} becomes the MMV \eqref{eq:MMV}.

\vspace{-7mm}
\subsection{Geometric Characterization}\label{sec:GLM_geo}
\vspace{-2mm}
In this section, we extend the characterization in Section \ref{sec:MMV_MMR} for linear regression to the GLM-based MMR problem. 
To facilitate our discussion, 
we assume that $\bbP^{(1)},\cdots,\bbP^{(K)}$ share a common covariate distribution, and hence a common distance-generating function $\sfA_{1}=\cdots=\sfA_{K}\equiv\sfA$. 
As a consequence, the equivalent min-max-distance problem in Corollary \ref{coro:MMR_GLM} is based on a common Bregman divergence $\sfD_{\sfA}(\cdot\|\cdot)$ across training $K$ populations.

The min-max-distance problem in Corollary \ref{coro:MMR_GLM} can be written in two equivalent forms:
\begin{align}
	\label{eq:GLM_primal}
	(\text{primal}) &&
	\min_{\theta\in\bbR^{p},R \ge 0} 
	&\Big\{R ~ \text{subject to} ~ \sfD_{\sfA}(\beta_{k}\|\theta) \le R, ~ 1 \le k \le K\Big\}; \\
	\label{eq:GLM_conjugate}
	(\text{conjugate}) &&
	\min_{\theta^{*}\in\bbR^{p},R \ge 0}
	&\Big\{R ~ \text{subject to} ~ \sfD_{\sfA^{*}}(\theta^{*}\|\mu_{k}) \le R, ~ 1 \le k \le K\Big\}.
\end{align}
The one-to-one correspondence between the primal solution $\theta^{\star}$ to \eqref{eq:GLM_primal} and the conjugate solution $\theta^{*\star}$ to \eqref{eq:GLM_conjugate} is $\theta^{*\star} = \nabla\sfA(\theta^{\star})$ $\Leftrightarrow$ $\theta^{\star} = \nabla\sfA^{*}(\theta^{*\star})$.
For interpretation,
the conjugate solution $\theta^{*\star} = \bbE\{ \bX\cA'(\bX^{\intercal}\theta^{\star}) \}$ is the covariate-prediction covariance, where $\cA'(\bX^{\intercal}\theta^{\star})$ is the MMR prediction of $Y$.
The geometry of the primal \eqref{eq:GLM_primal} and conjugate \eqref{eq:GLM_conjugate} problems are illustrated in Figure \ref{fig:MMR_GLM}.
In particular, the conjugate problem \eqref{eq:GLM_conjugate} aims to find the \textbf{minimal Bregman ball} $\cE_{\sfA^{*}}(\theta^{*\star},R^{\star}) = \{ \mu\in\bbR^{p}: \sfD_{\sfA^{*}}(\theta^{*\star}\|\mu) \le R^{\star} \}$ \textbf{enclosing} $\{\mu_{k}\}_{k=1}^{K}$, with $\theta^{*\star}$ and $R^{\star}$ as the optimized centroid and radius, respectively,
which are characterized below.

\begin{figure}[!t]
	\centering
	\includegraphics[width=0.49\textwidth]{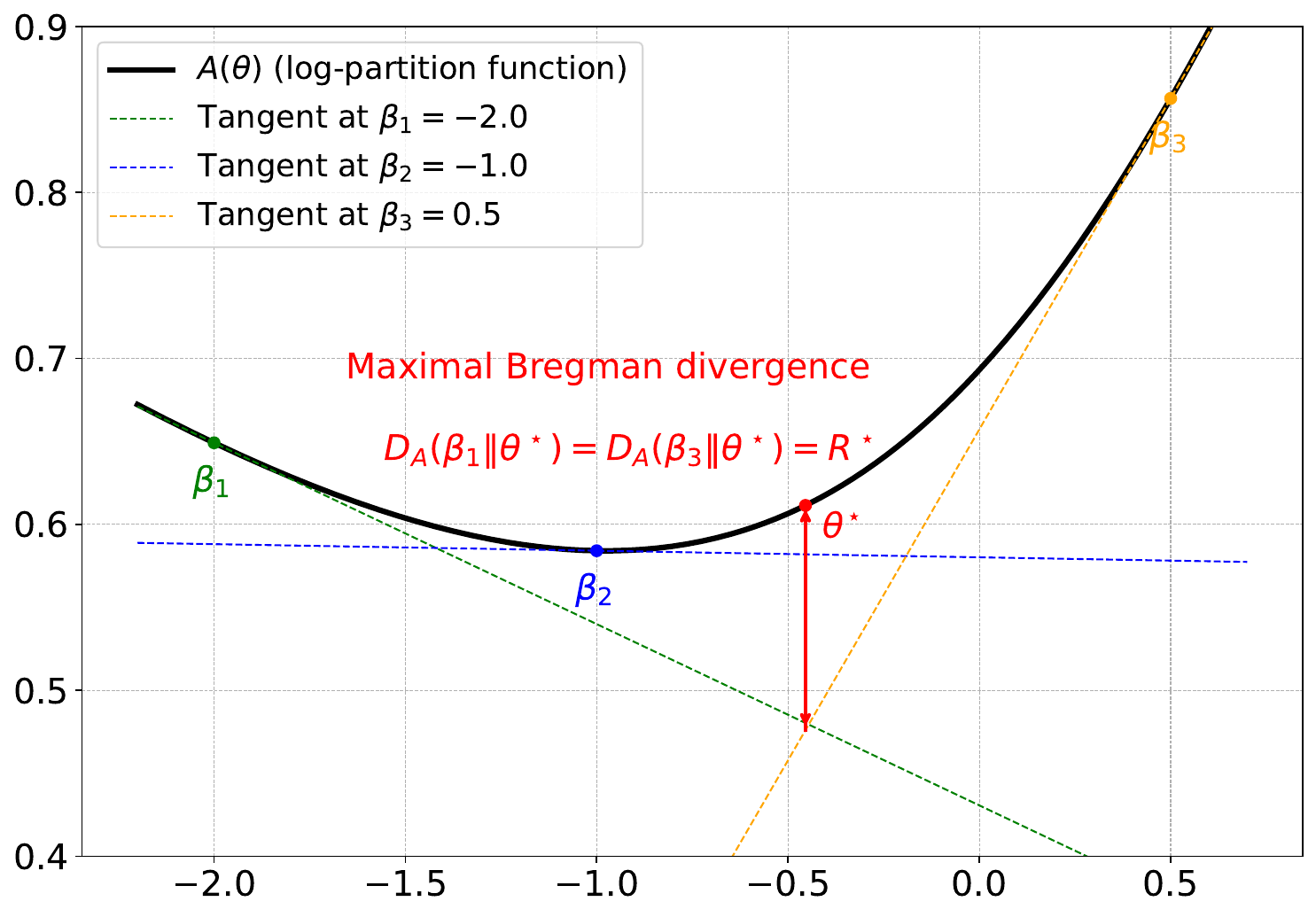}
	\includegraphics[width=0.49\textwidth]{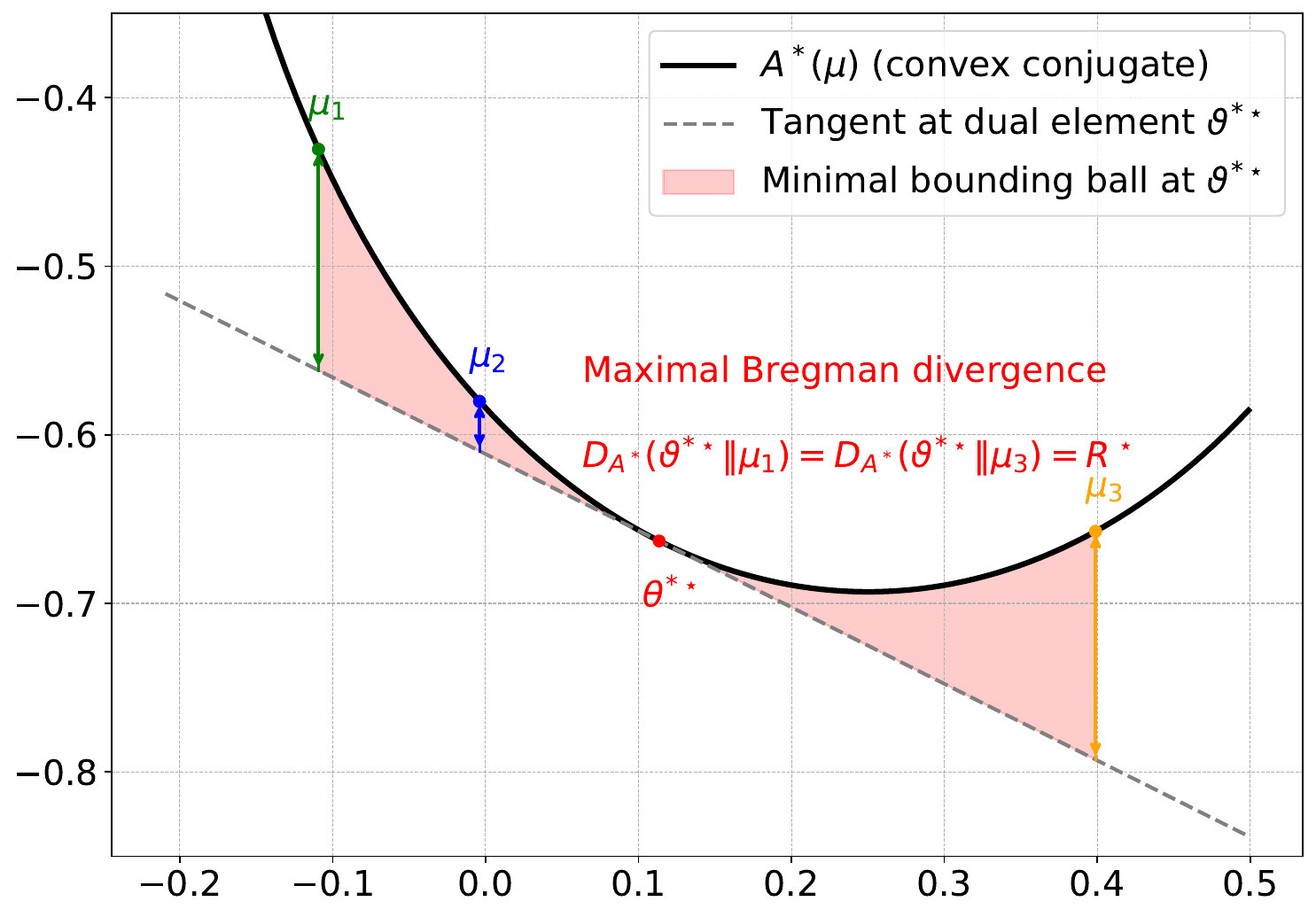}
	\caption{\footnotesize
		Geometric interpretation of MMR for logistic regression with three population regression coefficients $\beta_{1},\beta_{2},\beta_{3}\in\bbR$.
		\textbf{Left panel}:  the distance-generating function $\sfA(\theta)=\bbE_{\bbP}\log(1+e^{X\theta})$ with respect to $\theta\in\bbR$, with the primal coefficients $\beta_{1},\beta_{2},\beta_{3}$ marked in green, blue, and yellow. The Bregman divergence $\sfD_{\sfA}(\beta_{k}\|\theta)$  is the vertical distance between $\sfA(\theta)$ and the tangent line $\theta\mapsto\sfA(\beta_{k})+\nabla\sfA(\beta_{k})(\theta-\beta_{k})$ at $\beta_{k}$. 
		The primal solution $\theta^{\star}$ to \eqref{eq:GLM_primal} is where two tangent lines (green from $\beta_{1}$ and yellow from $\beta_{3}$) intersect, and the maximal vertical distance $R^{\star}$ (in red) identifies the supporting coefficients $\beta_{1}$ and $\beta_{3}$.
		\textbf{Right panel}: the conjugate distance-generating function $\sfA^{*}(\mu)$ with respect to $\mu\in\bbR$, with
	 the conjugate coefficients $\mu_{1},\mu_{2},\mu_{3}$ marked in green, blue, and yellow. 
		The conjugate Bregman divergence $\sfD_{\sfA^{*}}(\theta^{*\star}\|\mu_{k})$ is the vertical distance between $\sfA^{*}(\mu_{k})$ and the tangent line $\mu\mapsto\sfA^{*}(\theta^{*\star}) + \nabla\sfA^{*}(\theta^{*\star})(\mu-\theta^{*\star})$ at $\mu_{k}$.
		The conjugate solution $\theta^{*\star}$ to \eqref{eq:GLM_conjugate} is where the maximal distances on either side (green for $\mu_{1}$ and yellow for $\mu_{3}$) 
		are identical as~$R^{\star}$, corresponding to the supporting conjugate coefficients $\mu_{1}$ and $\mu_{3}$.
		The horizontal range from $\mu_{1}$ (in green) to $\mu_{3}$ (in yellow) forms the minimal Bregman ball $\cE_{\sfA^{*}}(\theta^{*\star},R^{\star})$ centered at $\theta^{*\star}$ and enclosing all  $\mu_{1},\mu_{2},\mu_{3}$.
	}
	\label{fig:MMR_GLM}
    \vspace{-4mm}
\end{figure}

\vspace{-2mm}
\begin{thm}[Characterization of GLM-Based MMR]\label{thm:MMR_GLM_dual}
	Consider the GLM risk in \eqref{eq:GLM}.
	Assume that $\bbP^{(1)},\cdots,\bbP^{(K)}$ satisfy Assumption \ref{asm:reg_GLM} and share a common covariate distribution, corresponding to a common distance-generating function $\sfA_{1}=\cdots=\sfA_{K}\equiv\sfA$.
	Then the ex-post MMR solution to \eqref{eq:regret_post} is  
	$\theta^{\star} = \nabla\sfA^{*}(\theta^{*\star})$,
	where
	$\theta^{*\star}=\sum_{k=1}^{K}\gamma_{k}^{\star}\mu_{k}$,
	and
	\begin{align}
		\gamma^{\star}\in 
		\argmax_{\gamma\in\Delta^{K-1}}\left\{
		\fR(\gamma)
		= \sum_{k=1}^{K}\gamma_{k}\sfA^{*}(\mu_{k}) - \sfA^{*}\left( \sum_{k=1}^{K}\gamma_{k}\mu_{k} \right) \right\}. 
		\label{eq:MMR_GLM_dual}
	\end{align}
	The min-max regret is $\displaystyle
    R^{\star} = \min_{\theta\in\bbR^{p}}\max_{1 \le k \le K}R(\theta,\bbP^{(k)}) = \fR(\gamma^{\star})$, 
	and we have $\{\mu_{k}\}_{k=1}^{K} \subseteq \cE_{\sfA^{*}}(\theta^{*\star},R^{\star})$.
	For any data-distribution $\bbQ$ with the same covariate distribution as $\bbP^{(1)},\cdots,\bbP^{(K)}$ and $\mu(\bbQ) = \bbE_{\bbQ}(\bX Y)\in\cE_{\sfA^{*}}(\theta^{*\star},R^{\star})$, 
	we have 
	$R(\theta^{\star},\bbQ)\le R^{\star}$, with equality if and only if $\mu(\bbQ)\in\partial\cE_{\sfA^{*}}(\theta^{*\star},R^{\star})$.
    The supporting set satisfies $\cK^{\star}=
    \{k:\mu_{k}\in\partial\cE_{\sfA^{*}}(\theta^{*\star},R^{\star})\} \supseteq \{k:\gamma_{k}^{\star} > 0\}$.
\end{thm}
\vspace{-2mm}

Theorem \ref{thm:MMR_GLM_dual} suggests a convex aggregation relationship among the ex-post MMR solution $\theta^{\star}$ and the local GLM parameters $\beta_{1},\cdots,\beta_{K}$ in the conjugate space as below:
\[ \begin{array}{cccc}
	& \text{(local populations)} &  &\text{(aggregation)}  \\
	\text{(primal)} \quad & 
	\beta_{1},\cdots,\beta_{K} &&
	\theta^{\star}
	\\
	& \nabla\sfA(\cdot)\downarrow 
	&&
	\uparrow\nabla\sfA^{*}(\cdot) \\
	\text{(conjugate)} \quad &
	\mu_{1}, \cdots, \mu_{K} &
	\overset{\eqref{eq:MMR_GLM_dual}}{\Longrightarrow} &
	\theta^{*\star}=\sum_{k=1}^{K}\gamma_{k}^{\star}\mu_{k}
\end{array} \]
The \textbf{dual MMR problem} \eqref{eq:MMR_GLM_dual} solves a \textbf{robust Bregman ball} $\cE_{\sfA^{*}}(\theta^{*\star},R^{\star})$ in the conjugate space.
Its centroid $\theta^{*\star}$ is a convex aggregation of the covariate-response covariances $\mu_{k}=\bbE_{\bbP^{(k)}}(\bX Y)$ for $k=1,\cdots,K$, 
and only those on the boundary $\partial\cE_{\sfA^{*}}(\theta^{*\star},R^{\star})$ correspond to positive aggregation weights.
In terms of generalization, Theorem \ref{thm:MMR_GLM_dual} suggests that $R^{\star}$ is the regret upper bound of $\theta^{\star}$ on those data-distribution $\bbQ$, whose covariate-response covariance $\mu(\bbQ)$ is enclosed in the robust Bregman ball $\cE_{\sfA^{*}}(\theta^{*\star},R^{\star})$,
and the regret upper bound is tight whenever $\mu(\bbQ)$ lies on the boundary $\partial\cE_{\sfA^{*}}(\theta^{*},R^{\star})$.

For comparison, we also establish the characterization for the GLM-based GDRO in Supplementary Material F.3, 
which extends the characterization for MMV in linear regression in Section \ref{sec:MMV_MMR}.
In particular, the dual GDRO problem maximizes the WMR among the mixtures of $\{\bbP^{(k)}\}_{k=1}^{K}$, and could be sensitive to their heterogeneity among  the training populations $\{\bbP^{(k)}\}_{k=1}^{K}$. As an extremely conservative scenario, if $\cov_{\bbP^{(k^{\star})}}(\bX, Y) = 0$ for some $\bbP^{(k^{\star})}$, then the maximal WMR is attained, and the GDRO estimator degenerates to 0.
}

\vspace{-7mm}
\section{Theoretical Properties}\label{sec:theory}
\vspace{-2mm}

In this section, we consider the theoretical properties of $\widehat{\theta}$ as the empirical MMR estimate in \eqref{eq:MMR}. 
In Sections \ref{sec:theory_K-MMR} and \ref{sec:theory_P-MMR},
we establish the excess MMR guarantees
based on the ex-post \eqref{eq:regret_post} and ex-ante \eqref{eq:regret_ante} criteria, respectively.
Their specialization to the linear regression and {\new GLM} settings are further discussed in Supplementary Material C.

\vspace{-7mm}
\subsection{Ex-Post MMR Guarantee}\label{sec:theory_K-MMR}
\vspace{-2mm}

The ex-post MMR guarantee is based on the uniform large-sample concentration of the empirical regret $R(\theta,\bbP_{n_{k}}^{(k)})$ on the population regret $R(\theta,\bbP^{(k)})$ as $n_{k}\to\infty$ for $k=1,\cdots,K$.
Two sets of generic concentration conditions are considered. 
For $k=1,\cdots,K$, let $\beta_{k} \in \argmin_{\beta \in \Theta}R^{\dagger}(\beta,\bbP^{(k)})$ be the risk minimizer on $\bbP^{(k)}$ whenever it exists.

\vspace{-2mm}
\begin{cond}\label{cond:regret_error}
	For every $k=1,\cdots,K$, assume that $\beta_{k}$ exists, 
	and for $t \ge 0$, with $\bbP^{(k)}$-probability at least $1 - 2e^{-t}$, we~have
	\[ \sup_{\theta \in \Theta}\left| {1 \over n_{k}}\sum_{i=1}^{n_{k}}\left[\ell_{\theta}\big(Z_{i}^{(k)}\big)-\ell_{\beta_{k}}\big(Z_{i}^{(k)}\big)\right] - \bbE_{\bbP^{(k)}}\left[\ell_{\theta}(Z)-\ell_{\beta_{k}}(Z)\right] \right| \le r_{n_{k}}(t), \]
	for some deterministic $r_{n_{k}}(t)$ that depends on the sample size $n_{k}$ and the parameter $t$.
\end{cond}
\vspace{-5mm}
\begin{cond}[\textcite{bartlett2005local}]\label{cond:regret_error_local}
	For every $k=1,\cdots,K$, assume that $\beta_{k}$ exists,
	and for $t \ge 0$ and $\eta > 1$, with $\bbP^{(k)}$-probability at least $1 - 2e^{-t}$, uniformly for all $\theta \in \Theta$, we have
	\[ \begin{aligned}
		& \bbE_{\bbP^{(k)}}\left[\ell_{\theta}(Z)-\ell_{\beta_{k}}(Z)\right] \le {\eta \over \eta - 1}{1 \over n_{k}}\sum_{i=1}^{n_{k}}\left[\ell_{\theta}\big(Z_{i}^{(k)}\big)-\ell_{\beta_{k}}\big(Z_{i}^{(k)}\big)\right] + \eta r_{n_{k}}^{2}(t); \\
		& {1 \over n_{k}}\sum_{i=1}^{n_{k}}\left[\ell_{\theta}\big(Z_{i}^{(k)}\big)-\ell_{\beta_{k}}\big(Z_{i}^{(k)}\big)\right] \le {\eta + 1\over \eta}\bbE_{\bbP^{(k)}}\left[\ell_{\theta}(Z)-\ell_{\beta_{k}}(Z)\right] + \eta r_{n_{k}}^{2}(t),
	\end{aligned}  \]
	for some deterministic $r_{n_{k}}(t)$ that depends on the sample size $n_{k}$ and the parameter $t$.
\end{cond}
\vspace{-2mm}

Condition \ref{cond:regret_error} is often considered in parametric estimation problems,
and is established for GLM in Supplementary Material G.2.
More generally, 
if $\ell_{\theta}(Z)$ is bounded,
$\Theta$ has a finite VC-dimension $\fC < +\infty$, and 
the Lipschitzness is satisfied:
$|\ell_{\theta'}(Z)-\ell_{\theta}(Z)| \le L(Z)\|\theta' - \theta\|$ with $\bbE_{\bbP^{(k)}}[L^{2}(Z)] < +\infty$, then Condition \ref{cond:regret_error} is a standard concentration result \parencite{van2023weak} with $r_{n}(t) \lesssim \sqrt{\fC\log(n/\fC)+t \over n}$.\footnote{
For $\{a_{n}\}$ and $\{b_{n}\}$, we denote $a_{n}\lesssim b_{n}$ if $a_{n}\le Cb_{n}$ for some universal constant $C < +\infty$. 
}
Condition \ref{cond:regret_error_local} is motivated from the local Rademacher complexities \parencite{massart2000some,bartlett2005local} based on the \textit{variance-expectation condition}: $\vvar_{\bbP^{(k)}}[\ell_{\theta}(Z) - \ell_{\beta_{k}}(Z)] \lesssim \bbE_{\bbP^{(k)}}[\ell_{\theta}(Z) - \ell_{\beta_{k}}(Z)]$.
One sufficient condition is the above Lipschitzness 
and the local strong convexity of the regret: $R(\theta,\bbP^{(k)}) \gtrsim \|\theta-\beta_{k}\|_{2}^{2}$ for $\theta$ in a neighborhood of $\beta_{k}$.

Based on either of Conditions \ref{cond:regret_error} and \ref{cond:regret_error_local}, we establish the excess ex-post MMR guarantee, which is based on the ``union bound'' of the concentration of $K$ empirical regrets. 

\vspace{-2mm}
\begin{thm}[Excess Ex-Post MMR Guarantee]\label{thm:excess_K-MMR}
	Consider the empirical MMR estimate $\widehat{\theta}$ in \eqref{eq:MMR}, 
	the ex-post MMR criterion $\cR_{\post}(\cdot)$ in \eqref{eq:regret_post},
	and $\cR_{\post}^{\star} = \min_{\theta \in \Theta}\cR_{\post}(\theta)$.
	Assume either of Conditions \ref{cond:regret_error} and \ref{cond:regret_error_local}.
	Let $r_{\post}(t) = \max_{1 \le k \le K}r_{n_{k}}(t + \log K)$.
	Then for every $t \ge 0$, with probability at least $1 - 2e^{-t}$, we have
	\[ \cR_{\post}(\widehat{\theta}) - \cR_{\post}^{\star} \le \begin{cases}
		4r_{\post}(t), 
		& \text{under Condition \ref{cond:regret_error}};\\
		4\sqrt{3\cR_{\post}^{\star}}r_{\post}(t) + (4+2\sqrt{6})r_{\post}^{2}(t), 
		& \text{under Condition \ref{cond:regret_error_local}}.
	\end{cases} \]
\end{thm}
\vspace{-2mm}

Let $n_{\min} = \min_{1\le k \le K}n_{k}$.
If Condition \ref{cond:regret_error} holds with $r_{n}(t) \lesssim \sqrt{\fC\log(n/\fC)+t \over n}$, then the excess ex-post MMR guarantee is $\cO_{\bbP}\big(\sqrt{\fC\log(n_{\min}/\fC)+t+\log K \over n_{\min}}\big)$.
If Condition \ref{cond:regret_error_local} holds with the same $r_{n}(t)$, and the risk minimizers are \textbf{homogeneous}: $\cR_{\post}^{\star} = 0\Leftrightarrow \beta_{1} = \cdots = \beta_{K}$,
then we can obtain a ``fast rate'' of the excess ex-post MMR guarantee $\cO_{\bbP}\big({\fC\log(n_{\min}/\fC)+t+\log K \over n_{\min}}\big)$.
A similar phenomenon of ``fast rate'' was pointed out by \parencite{bartlett2005local,agarwal2022minimax}.

\vspace{-7mm}
\subsection{Ex-Ante MMR Guarantee}\label{sec:theory_P-MMR}
\vspace{-2mm}

The ex-ante MMR guarantee is built upon the large-sample concentration in Section \ref{sec:theory_K-MMR}, and an additional large-$K$ concentration of the empirical maximum $\cR_{\post}(\theta)=\max_{1\le k \le K}R(\theta,\bbP^{(k)})$ on the essential supremum $\cR_{\ante}(\theta) = \sup_{\bbP\in\cP}R(\theta,\bbP)$.
In particular, we consider $R(\theta,\bbP)$ as a stochastic function in $\theta\in\Theta$ where $\bbP$ is ex-ante random under the meta-distribution $\fM_{\train}$.
For a fixed $\theta\in\Theta$, we call the real-valued random variable $R(\theta,\bbP)$ as the \textit{regret profile}.
We first consider a fixed-$\theta$ concentration condition for the empirical maximum $\cR_{\post}(\theta) = \max_{1 \le k \le K}R(\theta,\bbP^{(k)})$, 
where $\{\bbP^{(k)}\}_{k=1}^{K}\overset{\IID}{\sim}\fM_{\train}$.
Recall that $\cP=\supp(\fM_{\train})$.

\vspace{-2mm}
\begin{cond}[Locally Sub-Weibull Regret Profile]\label{cond:sub-Weibull}
	There exists some universal parameters $\alpha,t_{0} > 0$ and $\nu < +\infty$, such that for every fixed $\theta \in \Theta$, 
	the regret profile $R(\theta,\bbP)$ has a finite essential supremum $\cR_{\ante}(\theta) = \sup_{\bbQ\in\cP}R(\theta,\bbQ) < +\infty$ under $\bbP\sim\fM_{\train}$.
	Moreover, 
	\[ \fM_{\train}\Big\{ \cR_{\ante}(\theta) - R(\theta,\bbP) \le \nu t \Big\} \ge 1 - e^{-t^{\alpha}}; \quad \forall 0 \le t \le t_{0}. \]
\end{cond}
\vspace{-2mm}

Condition \ref{cond:sub-Weibull} is imposed on the meta-distribution of the regret profile $R(\theta,\bbP)$ around its essential supremum of $R(\theta,\bbP)$.
It is motivated from the regularity conditions based on which the extreme value theorem \parencite{haan2006extreme} holds: 
for fixed $\theta\in\Theta$,
$K^{1/\alpha}\{\cR_{\rm ante}(\theta) - \cR_{\rm post}(\theta)\}$ weakly converges to a Weibull-type extreme value distribution as $K\to\infty$.
Here, $\alpha$ is the extreme value index that determines the rate of convergence $\cR_{\ante}(\theta) - \cR_{\post}(\theta) = \cO_{\bbP}(K^{-1/\alpha})$.
We further require that the parameters $\alpha,\nu,t_{0}$ are universal across all $\theta \in \Theta$, which is necessary for such a weak convergence to be simultaneous.

Next, we consider an additional condition for the uniform concentration of $\cR_{\post}(\theta)$ as a stochastic function in $\theta\in\Theta$ under $\fM_{\train}$.

\vspace{-2mm}
\begin{cond}[Regret Function Complexity]\label{cond:parametric_entropy}
	Let $\sR=\{R(\theta,\bbP): \theta\in\Theta \}$.
	Assume that there exists a universal constant $0 < A < +\infty$ and a finite $\fC < +\infty$, such that 
	$\log \cN_{[]}(r,\sR,\cL^{\infty}(\fM_{\train})) \le \fC\log{A \over r}$ for $0 \le r \le A$.
	Here, $\cN_{[]}(r,\sR,\cL^{\infty}(\fM_{\train}))$ is the minimum number of $r$-brackets in $\cL^{\infty}(\fM_{\train})$ to cover $\sR$ \parencite{van2023weak}.
\end{cond}
\vspace{-2mm}

Condition \ref{cond:parametric_entropy} ensures that a uniform concentration of $\{\cR_{\post}(\theta):\theta\in\Theta\}$ can be established.
One sufficient condition is: $\Theta$ has a finite VC-dimension $\fC<+\infty$ and
$|\ell_{\theta'}(Z)-\ell_{\theta}(Z)| \le L(Z)\|\theta' - \theta\|$ with $\sup_{\bbP\in\cP}\bbE_{\bbP}|L(Z)| < +\infty$.
With Conditions \ref{cond:sub-Weibull} and \ref{cond:parametric_entropy} in addition to either of Conditions \ref{cond:regret_error} and \ref{cond:regret_error_local}, we are able to establish the excess ex-ante MMR guarantee. 

\vspace{-2mm}
\begin{thm}[Excess Ex-Ante MMR Guarantee]\label{thm:excess_P-MMR}
	Consider the empirical MMR estimate $\widehat{\theta}$ in \eqref{eq:MMR}, 
	the ex-ante MMR criterion $\cR_{\ante}(\cdot)$ in \eqref{eq:regret_ante},
	and $\cR_{\ante}^{\star} = \min_{\theta \in \Theta}\cR_{\ante}(\theta)$.
	Assume Conditions \ref{cond:sub-Weibull}, \ref{cond:parametric_entropy} and either of Conditions \ref{cond:regret_error} and \ref{cond:regret_error_local}.
	Let $r_{\post}(t) = \max_{1 \le k \le K}r_{n_{k}}(t + \log K)$ and
	$r_{\ante}(t) = (\max\{1,\alpha^{-1/\alpha}\}+A)\nu\left( {\fC\log(K/\fC) + \fC\log(1/\nu) + t \over K} \right)^{1/\alpha}$.
	Then for every $K,t\ge 0$ satisfying ${\fC \over K}\big({1 \over \alpha}\log{K \over \fC} + \log{1 \over \nu}\big) \le t_{0}^{\alpha}/2$ and $t \le t_{0}^{\alpha}K/2$, with probability at least $1 - 3e^{-t}$, we have
	\[ \cR_{\ante}(\widehat{\theta}) - \cR_{\ante}^{\star} \le \begin{cases}
		4r_{\post}(t) + r_{\ante}(t), & \text{under Condition \ref{cond:regret_error}}; \\
		4\sqrt{3\cR_{\ante}^{\star}}r_{\post}(t) + (4+2\sqrt{6})r_{\post}^{2}(t) + r_{\ante}(t), & \text{under Condition \ref{cond:regret_error_local}}.
	\end{cases} \]
\end{thm}
\vspace{-2mm}

Compared to Theorem \ref{thm:excess_K-MMR} for ex-post MMR, Theorem \ref{thm:excess_P-MMR} for ex-ante MMR 
incorporates an additional generalization error $r_{\ante}(t)$ due to the concentration of $\cR_{\post}(\cdot)$ on $\cR_{\ante}(\cdot)$. 
If Condition \ref{cond:regret_error} or \ref{cond:regret_error_local} holds with $r_{n}(t) \lesssim \sqrt{\fC\log(n/\fC)+t \over n}$,
then the excess ex-ante MMR guarantee becomes $\cO_{\bbP}\Big(\sqrt{\fC\log(n_{\min}/\fC)+\log K \over n_{\min}} + \big({\fC\log (K/\fC) \over K}\big)^{1/\alpha}\Big)$. 
Analogously to Theorem \ref{thm:excess_K-MMR}, 
it can be improved to a ``fast rate'' $\cO_{\bbP}\Big({\fC\log(n_{\min}/\fC)+\log K \over n_{\min}} + \big({\fC\log (K/\fC) \over K}\big)^{1/\alpha}\Big)$ under the ex-ante homogeneity $\cR_{\ante}^{\star} = 0$, in which case there exists a common risk minimizer shared across $\cP$.
\vspace{-7mm}
\section{Simulation Studies}\label{sec:simulation}
\vspace{-2mm}
In this section, we compare the MMR with the pooled ERM, GDRO, and MMV on their generalizability and sensitivity to various heterogeneity through extensive simulation studies, {\new including both multiple linear and logistic regression.  
Additional results and details, including the validation of the guarantees in Section \ref{sec:theory}, are provided in Supplementary Material~I.}

\vspace{-7mm}
\subsection{Multiple Linear Regression}
\vspace{-2mm}
\label{subsec: simu lm}
We first compare the four methods in multiple linear regression and investigate their generalizability under varying meta-distributions and their sensitivity to different degrees of heterogeneity in WUVs and WEVs across populations.

\noindent \textbf{Data Generation and Evaluation} \quad 
Following the hierarchical model in Section~\ref{sec:generalization}, 
we consider a meta-distribution that generates the regression coefficient $\beta\in\bbR^{p}$,
and a data-distribution that generates $(\bX,Y)\in\bbR^{p}\times\bbR$ satisfying $Y = \bX^{\intercal}\beta + \epsilon$, $\bX\indep\epsilon$ under a well-specified linear regression model.\footnote{
    Our method does not rely on a well-specified model. 
    We consider a well-specified model in simulation so that we can directly specify the regression coefficient, WUV, WEV for every data-distribution and study the impacts of their heterogeneity across training data-distributions.
}

For the training meta-distribution, $\beta$ is generated from a mixture of two uniform distributions $\fM_{\train}=\pi\mathsf{Uniform}(\sfB_{1})+(1-\pi)\mathsf{Uniform}(\sfB_{0})$ characterized by the mixture weight $\pi\in[0,1]$,
where 
$\sfB_{1}$ is the ball centered at $(3,3,\cdots,3)^{\top}\in\bbR^{p}$ with radius $3$,
$\sfB_{0}$ is the ball centered at $(1, 3,3,\cdots,3)^\top\in\bbR^{p}$ with radius $1$, and $\sfB_{0} \subseteq \sfB_{1}$.
In particular, we generate the regression coefficients $\{\beta_{k}\}_{k=1}^{K}\overset{\IID}{\sim}\fM_{\train}$ for $K=100$ populations. 
For $k=1,\cdots,K$,
we further generate the sample $\cD^{(k)}=\big\{\bX_{i}^{(k)},Y_{i}^{(k)}\big\}_{i=1}^{n_{k}}$ independently,
where
$\bX_{i}^{(k)}\sim\cN_{p}(0,\Ib_{p})$, $\epsilon_i^{(k)} \sim \mathcal{N}(0, p + \sigma^2\|\beta_k\|_2^2)$, $\bX_{i}^{(k)} \indep \epsilon_{i}^{(k)}$, and $Y_{i}^{(k)}={\boldsymbol{X}_{i}^{(k)\intercal}} \beta_k+\epsilon_i^{(k)}.$
Here, 
both the WEV $\nu_{k}^{2} = \|\beta_{k}\|_{2}^{2}$ and
the WUV $\sigma_{k}^{2} = p + \sigma^{2}\|\beta_{k}\|_{2}^{2}$ are heterogeneous across $k=1,\cdots,K$.
and a larger $\sigma^{2}$ induces a higher heterogeneity among the WUVs. 
By default, we set $\pi=0.2$, $\sigma^2=0.5$, $p=5$, and an equal sample size $n_k=1,000$ for $k=1,\cdots,K$.

To assess the generalizability to unseen testing populations, we evaluate the ex-ante worst-case regret \eqref{eq:regret_ante} over all ex-ante realizable populations. 
For every $\theta\in\bbR^{p}$, the ex-ante worst-case regret is computed as $\cR_{\ante}(\theta) := \sup_{\beta\in\sfB_1}\|\theta-\beta\|^2_2$, where  $\sfB_{1}$ contains all ex-ante realizable regression coefficients in the above data generating process.\footnote{
    In Supplemental Material~I,
    we provide additional results for other evaluation metrics, including the 
    ex-ante expected and worst-case risks,
    and ex-ante worst-case explained variance. 
    We also provide visualizations for the estimators of four methods.
}
 
\begin{figure}[!h]
\centering
\includegraphics[width=\linewidth]{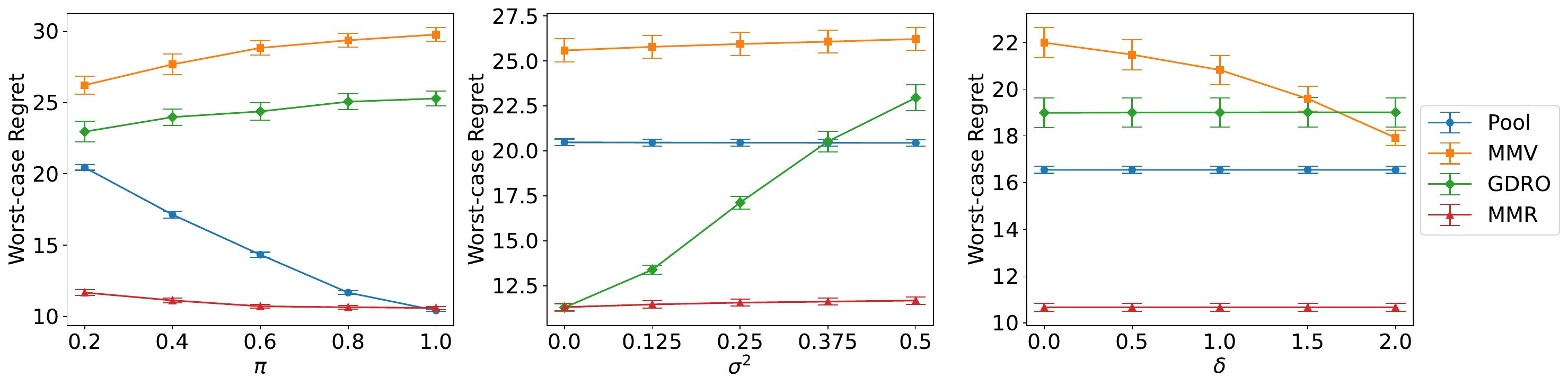}
\caption{The worst-case regret of four empirical estimates under multiple linear regression, with error bars indicating standard errors across 30 independent data-generating replications. From left to right: the results under meta-distributional shift, varying degrees of heterogeneity in within-population unexplained variance, and within-population explained variance.}
\label{fig:simulation-main-new}
\vspace{-4mm}
\end{figure}

\noindent \textbf{Generalization Robustness} \quad
To assess the generalization robustness,
we vary $\fM_{\train}$ through the mixture weight $\pi$ in $\{0.2,0.4,0.6,0.8,1.0\}$.
As $\pi$ decreases, we observe more training regression coefficients from $\sfB_{0}$ as a restricted subset of $\sfB_{1}$. 
In the left panel of Figure~\ref{fig:simulation-main-new}, the pooled estimator is not robust against the variation of training meta-distribution. 
In particular, for $\pi=1$, the pooled estimator shares a similar performance as the MMR estimator. 
As $\pi$ decreases, the pooled estimator experiences a substantial increase in the worst-case regret. In contrast, the MMV, GDRO and MMR estimators perform stably as $\pi$ varies, which suggests their generalization robustness. 
In particular, MMR enjoys the best ex-ante worst-case regret guarantee.

\noindent \textbf{Sensitivity to the Heterogeneity in WUVs} \quad 
We vary the degree of heterogeneity in WUVs through $\sigma^2$ in $\{0, 0.125, 0.25, 0.375, 0.5\}$. 
When $\sigma^{2} = 0$, the WUVs $\sigma_{k}^{2}=p$ for $k=1\cdots,K$ are homogeneous. As $\sigma^{2}$ increases, the WUVs $\{\sigma_{k}^{2}=p+\sigma^{2}\|\beta_{k}\|_{2}^{2}\}_{k=1}^{K}$ are increasingly heterogeneous, 
and up to a sufficiently large $\sigma^{2}$, the population $k^{\star}\in\argmax_{1\le k \le K}\|\beta_{k}\|_{2}^{2}$ dominates as in Table \ref{tab:criteria}, leading to a degenerate GDRO estimator.
In the middle panel of Figure~\ref{fig:simulation-main-new}, the pooled, MMV, and MMR estimators remain stable as $\sigma^{2}$ varies,
while the GDRO estimator becomes worse as $\sigma^{2}$ increases. 
This suggests that GDRO is sensitive to the heterogeneous WUVs, and can suffer from degeneration when certain WUV dominates.

\noindent \textbf{Sensitivity to the Heterogeneity in WEVs} \quad
We also evaluate the four methods under varying degree of heterogeneity in WEVs. 
In particular, for $\delta$ in $\{0, 0.5, 1, 1.5, 2\}$,
we let $\vec{\delta}:=(\delta,\delta,\cdots,\delta)^{\top}$,
and shift the support of the training meta-distribution $\fM_{\train}$ toward the origin by replacing each $\beta \in \mathsf{B}_{1}$ with $\beta - \vec{\delta}$. 
Consequently, the training regression coefficients become $\beta_{k} - \vec{\delta}$ for $k=1,\ldots,K$.
Such a translation can affect the heterogeneity among the WEVs $\{\nu_{k}^{2}=\|\beta_{k}-\vec{\delta}\|_{2}^{2}\}_{k=1}^{K}$, while the worst-case regret remains unchanged.
As shown in the right panel of Figure \ref{fig:simulation-main-new}, the MMV estimator is sensitive to the variation of $\delta$ due to its dependency on the WEVs, while the pooled, GDRO and MMR estimators are insensitive to such a variation.

{\new 
\vspace{-7mm}
\subsection{Multiple Logistic Regression}
\vspace{-2mm}

We further compare the four methods in multiple logistic regression. 
In this setting, the pooled, GDRO and MMR estimators are based on the GLM risk in \eqref{eq:GLM}. 
We define MMV for logistic regression 
through the revised explained variance criterion from \eqref{eq:MMV} as $V(\theta,\bbP) = \vvar_{\bbP}(Y) - \bbE_{\bbP}[Y - \cA'(\bX^{\intercal}\theta)]^{2}$,
where $\cA'(\bX^{\intercal}\theta)$ is the GLM-based prediction, and for logistic regression, $\cA'(\eta)={e^{\eta} \over 1 + e^{\eta}}$ for $\eta\in\bbR$.

\noindent \textbf{Data Generation and Evaluation} \quad 
We follow the data generating process analogous to Section \ref{subsec: simu lm} but under a well-specified logistic regression model. 
In the first stage, we generate the training regression coefficients $\{\beta_{k}\}_{k=1}^{K}$ independently from $\fM_{\train} = \pi\mathsf{Uniform}(\sfB_{1}) + (1-\pi)\mathsf{Uniform}(\sfB_{0})$, where $\sfB_{0}$ is the ball centered at $(3,3)\in \bbR^2$ with radius $3$ and $\sfB_{1}$ is the ball centered at $(1,3)\in \bbR^2$ with radius $1$.
In the second stage, for $k=1,\cdots,K$,
we generate the sample $\cD^{(k)} = \big\{\bX_{i}^{(k)}, Y_i^{(k)}\big\}_{i=1}^{n_k}$ independently,
where 
$\bX_{i}^{(k)} \sim \cN_{p}(0.5,\Ib_{p})$, 
and $Y_i^{(k)}\big|\bX_{i}^{(k)} \sim \mathrm{Bernoulli}\left(\cA'(\boldsymbol{X}_i^{(k) \top} \beta_k)\right)$. 
As in Section \ref{subsec: simu lm}, our primary evaluation metric is the ex-ante worst-case regret $\cR_{\ante}(\theta) = \sup_{\beta\in\sfB_{1}\cup\sfB_{0}}\sfD_{\sfA}(\beta\|\theta)$ based on Proposition \ref{prop:GLM}. Details on the four methods and evaluation metrics are provided in Supplementary Material~I.
\begin{figure}[!ht]
	\centering 
    \vspace{-2mm}
	\includegraphics[width=0.9\textwidth]{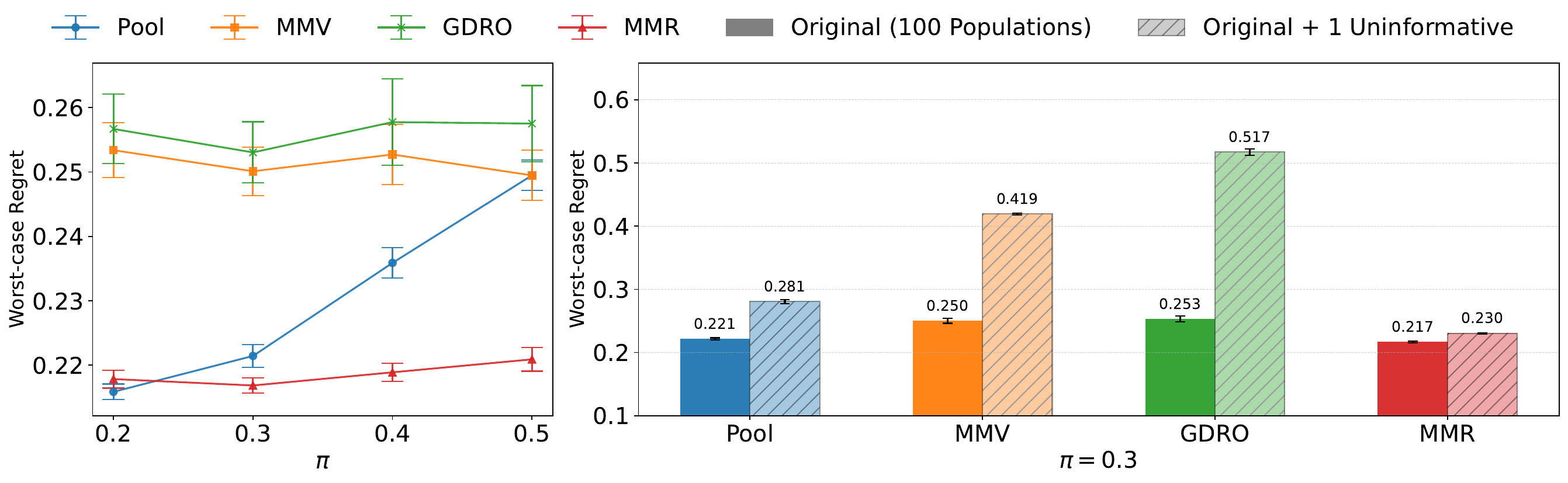}
	\caption{Worst-case regret of four empirical estimators under multiple logistic regression, with error bars indicating standard errors across 30 independent data-generating replications. 
	Left: the results under meta-distributional shift.  
	Right: the results when one additional uninformative sample is included. 
	}
    \vspace{-4mm}
	\label{fig:glm performance}
\end{figure}

\noindent \textbf{Generalization Robustness} \quad 
We vary $\fM_{\train}$ through the mixture weight $\pi$ in $\{0.2, 0.3, 0.4, 0.5\}$. 
As shown in the left panel of \Cref{fig:glm performance}, 
the pooled estimator is not robust against such a meta-distributional variation, while the robust methods MMV, GDRO and MMR perform stably.
In particular, MMR achieves the best ex-ante worst-case regret.

\noindent \textbf{Sensitivity to Uninformative Samples} \quad 
In Section \ref{sec:GLM_geo}, 
we point out an extremely conservative scenario when $\cov_{\bbP^{(k^{\star})}}(\bX, Y) = 0$ for some $k^{\star}$,
which corresponds to an \textbf{uninformative population} $\bbP^{(k^{\star})}$ since the logistic regression captures no relationship between $\bX$ and $Y$ on $\bbP^{(k^{\star})}$.
Such an uninformative population can dominate GDRO and MMV, resulting in a degenerate zero estimator.
Motivated by this observation, we simulate the practical scenario that the training meta dataset may contain some uninformative samples, leading to conservative GDRO and MMV estimates.
Specifically, we follow the same data generating process as above, 
and introduce one additional \textbf{uninformative sample} $\cD^{(K+1)}$, where $Y_i^{(K+1)} \sim \mathrm{Bernoulli}(0.5)$ independently of $\bX_{i}^{(K+1)}$.
Then we apply all methods to $\{\cD^{(k)}\}_{k=1}^{K+1}$.
As shown in the right panel of \Cref{fig:glm performance}, the performance of MMV and GDRO deteriorates substantially after involving the uninformative sample, which indicates their sensitivity and conservation in such a scenario.
In contrast, the pooled and our MMR estimators are less affected by the additional uninformative sample.
}

\vspace{-7mm}
\section{Real Data Examples}
\vspace{-2mm}
\label{sec:realdata}
In this section, we apply the MMR method to an image recognition task to demonstrate its generalizability to unseen populations. We also provide additional results for a regression problem on post-transplant mortality prediction in Supplementary Material~J.

Predictive models often face challenges when deployed across diverse and unseen populations different from those in the training data. To illustrate the effectiveness of our method in addressing such heterogeneity, we consider a facial attribute recognition task. 
We use the CelebA face image dataset~\parencite{liu2015faceattributes}, which contains celebrity images annotated with $40$ binary attributes. Following~\textcite{sagawa2019distributionally}, the target task is to classify the \texttt{Blond Hair} attribute. To create naturally heterogeneous populations, we partition the data according to the Cartesian product of four additional attributes: \texttt{Male}, \texttt{Young}, \texttt{Receding Hairline}, and \texttt{Blurry}. After removing groups with fewer than $500$ samples, we retain $12$ groups totaling $201,688$ images. We extract $128$-dimensional features for each image using a pre-trained ResNet-18~\parencite{he2016deep}, which provides a feature representation as input to a logistic regression classifier that predicts the presence of \texttt{Blond Hair}.

\begin{figure}[!t]
	\centering
\includegraphics[width=0.49\textwidth]{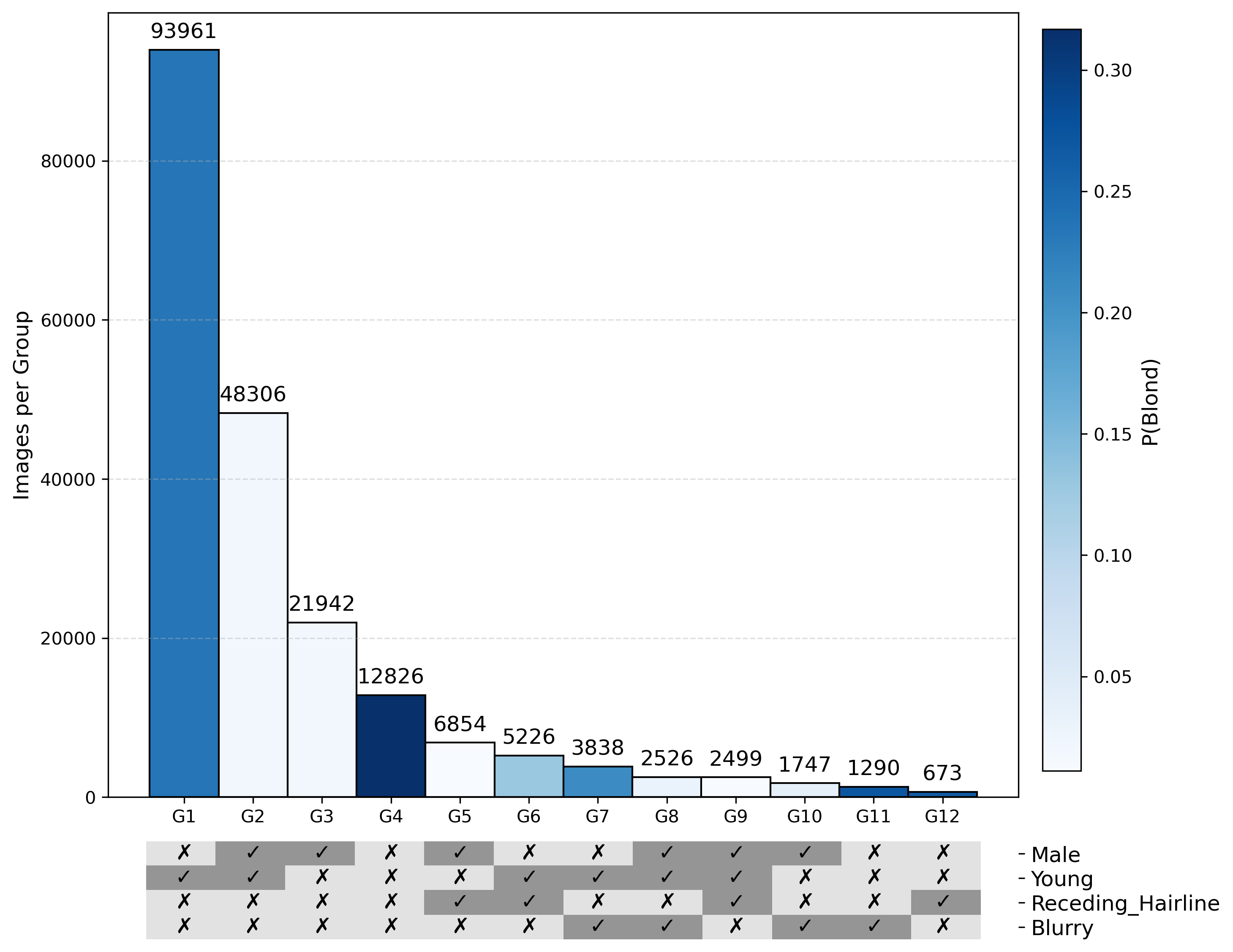}
	\includegraphics[width=0.45\textwidth]{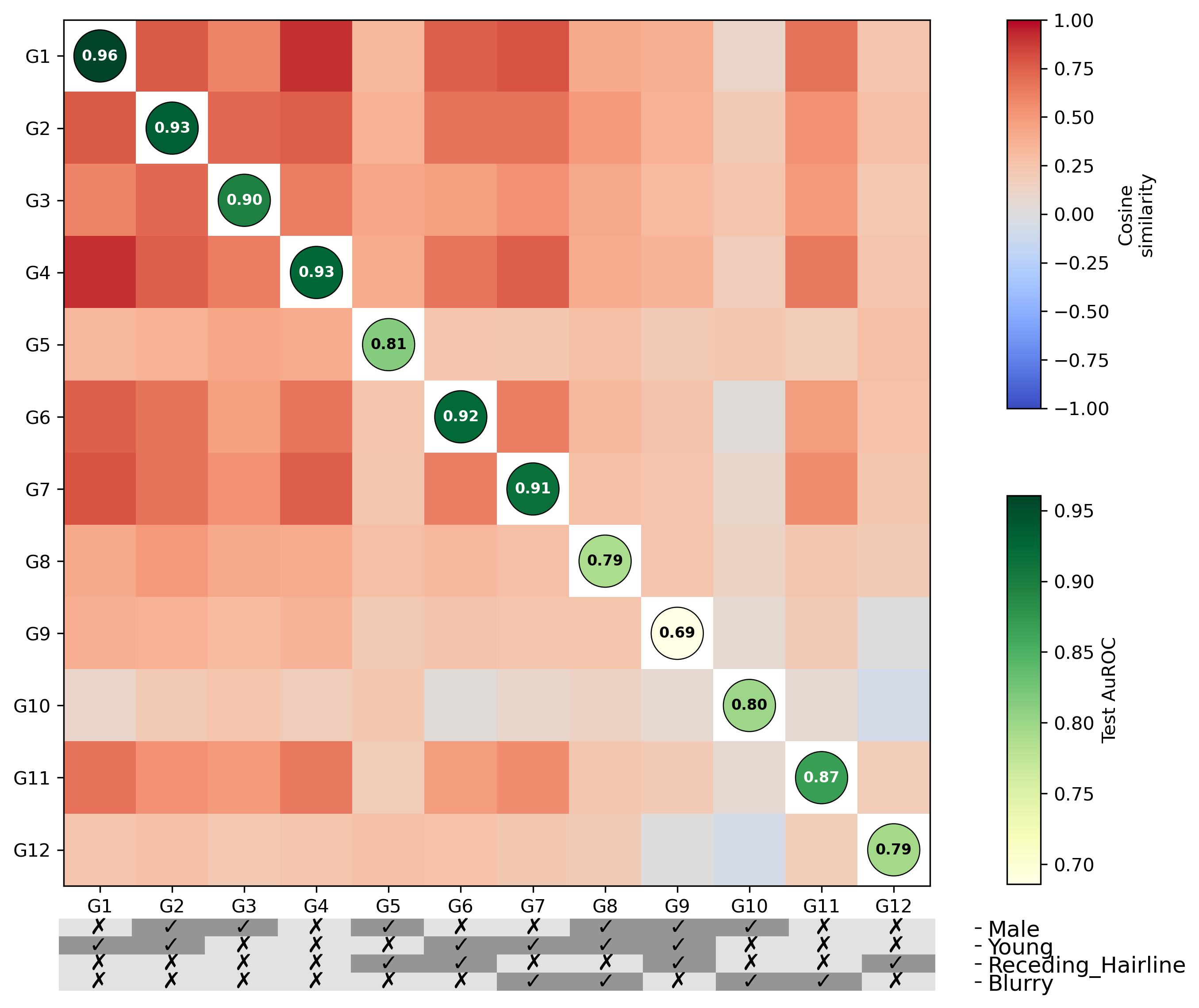}
	\caption{\textbf{Left:} Bar plot showing the sample size of each group, with groups defined by combinations of the attributes \texttt{Male}, \texttt{Young}, \texttt{Receding Hairline}, and \texttt{Blurry}. Each bar is colored by the proportion of positive labels for the target \texttt{Blond Hair}. 
\textbf{Right:} Heatmap of cosine similarity between local logistic regression parameters fitted separately within each group. Diagonal entries are the within-population testing AuROC.
}
    \vspace{-4mm}
	\label{fig:celebA data heterogeneity}
\end{figure}

As shown in the left panel of \Cref{fig:celebA data heterogeneity}, the dataset shows substantial imbalance in both sample sizes and label prevalence across groups. Specifically, the sample sizes range from $673$ to $93,961$, and the proportion of positive labels varies between 5\% and 30\%. To assess the heterogeneity in local GLM parameters and WMRs, we fit within-sample logistic regressions. The right panel of \Cref{fig:celebA data heterogeneity} shows the cosine similarity matrix of these fitted coefficients and reports the within-population testing Area Under the Receiver Operating Characteristic curve (AuROC) along the diagonal, which ranges from 0.69 to 0.98. These results suggest substantial heterogeneity across populations. Our goal is to develop a classifier trained on such heterogeneous populations that can generalize well to unseen ones.

\begin{wraptable}{o}{0.35\textwidth}
\captionsetup{font=footnotesize}
\vspace{-5mm}
\centering
\caption{Leave-one-group-out performance comparison (mean across all groups and 30 replications). Higher AuROC and lower Brier score indicate better performance.}
\label{tab: celeba leave one out}
\scalebox{0.8}{
\begin{tabular}{lcc}
\toprule
Method & AuROC & Brier \\
\midrule
Within-Population & 0.847 & 0.081 \\
Pooled ERM & 0.877 & 0.076 \\
MMV & 0.830 & 0.092 \\
GDRO & 0.831 & 0.105 \\
MMR (ours) & \textbf{0.890} & \textbf{0.061} \\
\bottomrule
\end{tabular}
}
\end{wraptable}

We compare the MMR estimator with the pooled ERM, GDRO, and MMV estimators. 
To evaluate generalization performance on unseen populations, we adopt a \textit{leave-one-group-out} strategy. Specifically, in each round, one group is held out entirely as a testing unseen population, and the model is trained on the remaining groups. Within each held-out group, we further split the data in a 1:1 ratio and use the training split to fit a baseline ``within-population'' estimator, providing a reference for models trained only on the data from that group.  All methods are then evaluated on the remaining test split of the held-out group. We report the average AuROC and Brier score over 30 random splits and all held-out groups.

The results in \Cref{tab: celeba leave one out} show that the proposed MMR achieves the best prediction performance in unseen groups, with the highest average AuROC and the lowest Brier score. 
In particular, MMR outperforms the within-population estimator, demonstrating better generalizability to unseen populations by leveraging diverse populations in the training data  than using data from the target group alone. 
The pooled estimator performs reasonably well but remains suboptimal, as it tends to favor large groups and fails to generalize when the data distribution of the held-out group differs substantially from the majority.  
Both GDRO and MMV underperform the within-population baseline. 
GDRO suffers due to the high heterogeneity in WMRs as indicated by the large variation of the within-population testing AuROC in \Cref{fig:celebA data heterogeneity}, while MMV underperforms possibly because its use of the square loss is less favored for binary classification and may yield less calibrated probabilities.

\vspace{-7mm}
\section{Summary}
\vspace{-2mm}
In this work, we have introduced a general MMR framework for 
the robust performance on heterogeneous training populations (ex-post MMR), and 
the generalization to an unseen testing population (ex-ante MMR).
Under the robust hierarchical model, the empirical MMR can achieve both goals simultaneously.
Compared to other robust learning methods that can be considered as general GDRO instances of \eqref{eq:GDRO}, MMR is insensitive to the heterogeneous WMRs among the training populations,
and can avoid the degeneration to the risk minimization on a single dominating population.
In the linear regression and GLM settings, we characterize MMR as a min-max-distance problem that solves a robust ball via duality.
Our simulation studies and real data applicaitons further confirm the robustness of MMR and its insensitivity to the heterogeneous WMRs.

There are several future directions to be explored based on the proposed framework. 
In this work, we mainly specialize MMR to the parametric estimation problems. It would be interesting to employ the MMR framework in other problem setups. For example, it could be extended to learning optimal treatment regimes in precision medicine and the development of robust policies in statistical decision making \parencite{mo2021learning}. 
In addition, our proposed algorithm mainly focuses on a smooth and strongly convex loss. 
Another interesting direction is to develop algorithms for non-smooth, non-convex losses, such as the zero-one loss in classification. These extensions would further broaden the applicability of the MMR framework.

\begingroup
\setstretch{1}
\setlength\bibitemsep{1.5\itemsep}
\AtNextBibliography{\small}
\printbibliography
\endgroup

\end{document}